\documentclass[letterpaper, 10 pt, conference]{ieeeconf}  

\IEEEoverridecommandlockouts                              
\usepackage{cite}
\usepackage{caption}

\usepackage{subcaption}

\usepackage{amsmath,amssymb,amsfonts,bm}
\usepackage{algorithmic}
\usepackage{graphicx}
\usepackage{textcomp}
\usepackage{xcolor}
\def\BibTeX{{\rm B\kern-.05em{\sc i\kern-.025em b}\kern-.08em
    T\kern-.1667em\lower.7ex\hbox{E}\kern-.125emX}}

\usepackage{amsthm}
\newtheorem{assumption}{Assumption}
\newtheorem{lemma}{Lemma}
\newtheorem{theorem}{Theorem}
\newcommand{\norm}[1]{\left\lVert#1\right\rVert}
\usepackage{comment}
\usepackage{stfloats}
\usepackage{multirow}

\usepackage[normalem]{ulem}

\newcommand{\rev}[1]{\textcolor{black}{#1}}
\newcommand{\LSv}[2]{#2}
\usepackage[color=yellow, textsize=small, textwidth=\marginparwidth, draft]{todonotes}   
\title{\LARGE \bf
Quaternion-based non-singular terminal sliding mode control\\ for a satellite-mounted space manipulator
}

\author{Jacopo Giordano and Angelo Cenedese
\thanks{J. Giordano and A. Cenedese are with the Department of Information Engineering, University of Padova, Italy
        {\tt\small \{jacopo.giordano, angelo.cenedese\}@unipd.it}}%
}

\begin{document}
\bstctlcite{IEEEexample:BSTcontrol}

\maketitle
\thispagestyle{empty}
\pagestyle{empty}


\begin{abstract}
In this paper, a robust control solution for a satellite equipped with a robotic manipulator is presented. 
First, the dynamic model of the system is derived based on quaternions to describe the evolution of the attitude of the base satellite. 
Then, a non-singular terminal sliding mode controller that employs quaternions for attitude control, is proposed for concurrently handling all the degrees of freedom of the space manipulator. 
Moreover, an additional adaptive term is embedded in the controller to estimate the  upper bounds of disturbances and uncertainties. 
The result is a resilient solution able to withstand  unmodelled dynamics and interactions.
Lyapunov theory is used to prove the stability of the controller and \LSv{numerical simulations are performed to verify its performances in multiple scenarios.}{numerical simulations allow assessing performance and fuel efficiency.}
\end{abstract}


\section{Introduction}

Satellites equipped with manipulators will play a primary role in the future of space activities: their use is envisaged for a wide range of operations that can be categorized as on-orbit servicing, active debris removal, and on-orbit assembly and manufacturing~\cite{surveyIntr}. The last two kinds of missions are particularly challenging due to the presence of a large degree of uncertainty, linked to the effects of an unknown or partially known grasped object.

To control a system in presence of uncertainties, two popular approaches are adaptive control and robust control.
In the former, the parameters of the controller are tuned during operations, while in the latter the presence of disturbances is explicitly taken into account in the control design. \cite{slotine1991applied}. 
In~\cite{adaptiveFlex} an adaptive controller is proposed for a space manipulator to better deal with the vibrations generated by the elasticity resulting from the motors and the disturbances related to their non-linear frictions. Instead, \cite{angularMomAdap} proposes an adaptive control solution that is able to compensate for the angular momentum accumulated by the Reaction Wheels (RWs) actuators. 
In general, the stability and performances of adaptive controllers are strictly related to those of the observer and the employed model. Indeed, unmodelled dynamics and disturbances can affect the behavior of the controller and compromise its performance.
In terms of robust control techniques, there are two main approaches: $H_{\infty}$ control and Sliding Mode Control (SMC). The first one is particularly popular in the linear case since it is supported by powerful theoretical results and allows to evaluate analytically the controlled system robustness and stability. 
However, in its non-linear formulation such advantages are lost and tuning is generally performed by resorting to trial-and-error approaches. 
As an example in this direction, an $H_{\infty}$ controller for a space manipulator is developed in~\cite{hinfContrFloat} for a free-floating scenario. 
%
%
Non-linear SMC has been also considered for controlling space manipulators. 
\rev{However, problems related to excessive control effort and oscillations around the reference values and singularity issues concerning either the definition of the sliding surface or of the system model representation itself} may arise and need to be addressed with particular attention.
\rev{In this respect, some noteworthy works are reported next.}
\rev{While in \cite{li2021simultaneous} the general problem of convergence and sliding surface non-singularity is discussed, in~\cite{shao2021nonsingular} a non-singular terminal SMC (NTSMC) is developed specifically for trajectory tracking of space manipulators.} In~\cite{smNTSMC} a NTSMC controller is coupled with a radial basis function neural network to estimate part of the unmodeled dynamics. 
An integral SMC is designed in~\cite{jia2020continuous} to control a space manipulator in presence of disturbances and possible faults. 
In both these solutions, the switching gains of the sliding controllers are adapted using an estimate of the uncertainties norm upper-bound in order to mitigate the oscillations around the reference.
\rev{Differently, fuzzy logic is considered for such purpose in~\cite{qingxuan2021adaptive} and in~\cite{xie2020motion}, where, in the latter, tuning is obtained via reinforcement learning.}

Despite this research effort, to the best of the authors' knowledge, none of the SMC-based controllers for space manipulators \rev{explicitly considers model issues by adopting a singularity-free attitude representation (e.g. unit quaternions)} for the base satellite (BS), which, instead, could potentially lead to unfeasible control inputs and 
to instability \rev{during error correction}.
To fill this gap, a novel NTSMC controller is proposed in this work, for simultaneously controlling the BS and the manipulator joints of a free-flying space manipulator. 
Specifically, two main contributions can be highlighted:\\
$i)$ A novel sliding variable is considered that employs the quaternion algebra to describe the attitude error of the BS.
The designed sliding variable is able to capture the $\mathbb{S}^3$ topology  and address the unwinding phenomenon associated with the quaternion double coverage. 
Inspired by~\cite{unwinding}, continuity on the attitude control input is obtained by introducing a discontinuity in the sliding variable. Indeed, the shortest path to the desired orientation is always selected by the controller. 
The resulting control system does not \rev{suffer from any model singularities and it can improve efficiency}, especially for large orientation corrections.\\
$ii)$ An adaptive strategy is proposed to estimate an upper-bound of the uncertainties affecting the space manipulator: it extends the approach of~\cite{smNTSMC} by handling each degree of freedom of the system independently and considering a state dependent upper-bound based on the system velocities.

\section{Assumptions and notation}
\label{sec:ass&not}
In this work, a BS equipped with a robotic arm is considered: 
the joints are assumed to be revolute and the flexibility of the system elements is supposed to be negligible. 
Hence, the system can be modeled as a chain of $n+1$ rigid bodies, where $n$ is the number of arm links. 
\LSv{In addition, we assume the system to be fully actuated, i.e. able to produce instantaneous accelerations w.r.t. all its degrees of freedom.}{} 
In this work a proximity operation scenario is evaluated. The target object, which is used as reference point, is supposed to be fixed in the inertial frame. This assumption is not too simplistic when considering close proximity operations that are much shorter than the orbital periods of the two bodies~\cite{surveyIntr}. In addition, no orbital or environmental disturbances are considered since they are deemed to be considerably smaller than those related to the dynamics of the actuators during such a time span.

\LSv{
\begin{figure}[t]
    \centering
    \includegraphics[width=0.46\textwidth]{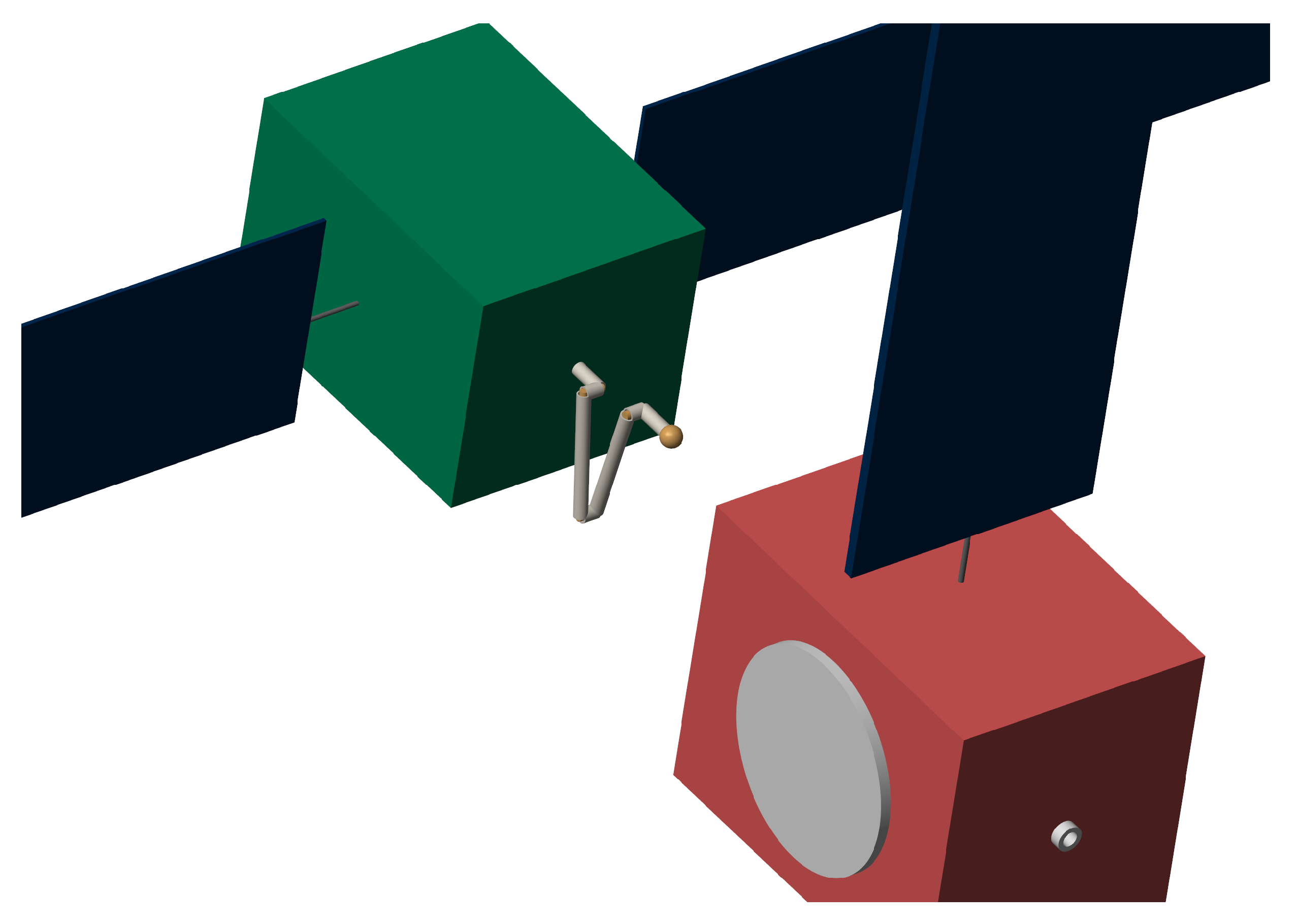}
    \caption{Space manipulator mounted on a BS (in green). The target is represented as the red object.}
    \label{fig:figSM}
\end{figure}
}{}

The considered frames are: an inertial frame $\mathcal{I}$, a body frame $\mathcal{B}$ rotating together with the satellite body and located at its Center of Mass (CoM), a frame $\mathcal{J}$ attached to the $j^{th}$ link of the robotic arm and placed at its CoM. 
%
The position vector of frame $\mathcal{K}$ w.r.t. to the $\mathcal{M}$ one expressed in frame $\mathcal{O}$ coordinates is represented with the symbol $\bm{p}_{mk}^o \in \mathbb{R}^3$; if the superscript or one of the subscripts are omitted, the vector is considered to be expressed in $\mathcal{I}$ coordinates or originated from such frame respectively. This notation applies also to linear and angular velocity vectors.
The generic rotation matrix describing the orientation of  $\mathcal{M}$ w.r.t. $\mathcal{K}$ is represented with the symbol $\bm{R}_{km} \in SO(3)$; again, the first frame is omitted if it coincides with $\mathcal{I}$. 
The same subscript notation is used for 
unit quaternions $\bm{q}_{km} = [\eta_{km} \ \bm{\epsilon}^T_{km}]^T \in \mathbb{S}^3$, where $\eta$ and $\bm{\epsilon}$ are its scalar and vectorial part respectively. The Hamiltonian product is represented with the symbol $\otimes$. 
A generic force or torque applied at the origin of  $\mathcal{M}$ and represented w.r.t. $\mathcal{O}$ is symbolized with $\bm{f}_{m}^o \in \mathbb{R}^3$ or $\bm{\tau}_{m}^o \in \mathbb{R}^3$ respectively; the superscript is not specified in the inertial frame case. 
The mass of the BS and that of the $j^{th}$ manipulator body are $m_{b}$ and $m_{j}$ respectively, whereas their inertia w.r.t. $\mathcal{I}$ are denoted with $\bm{I}_{b}$ and $\bm{I}_{j}$ respectively.
The angular positions of arm joints are symbolized with $\bm{q} \in \mathbb{R}^n$ and their torques are represented with $\bm{\tau} \in \mathbb{R}^n$. 
In general, if the subscript $d$ is added to a variable it symbolizes that such quantity is the desired reference value. The identity and zero matrices are $\bm{E}$ and $\bm{0}$ respectively, the skew symmetric matrix operator is symbolized by $[\cdot]_\times \in \mathbb{R}^{3  \times 3 }$ and the diagonal one by diag($\cdot$). The $\alpha \in \mathbb{R}$  power of a vector $\bm{v}= \left[v_1, \ldots,v_n \right]^\top\in \mathbb{R}^n$ is defined component-wise as follows: $\bm{v}^\alpha = \left[\vert v_1 \vert^\alpha \text{sgn}(v_1) , \ldots, \vert v_n \vert^\alpha \text{sgn}(v_n) \right]^\top$. Finally, the Euclidean norm is represented with the symbol $\norm{\cdot}$. 

\section{Dynamical model}
\label{sec:dynMod}
The system dynamics based on quaternions can be derived using the Lagrangian approach~\rev{(see \cite{wilde2018equations,surveyIntr} and references within for more details)}. \LSv{To this end the linear and angular velocities the $j^{th}$ arm link w.r.t. the inertial frame are found:
\begin{equation}
    \bm{v}_j = \bm{v}_b -[\bm{p}_{bj}]_\times \bm{\omega}_b + \bm{J}_{\bm{v}_j}(\bm{q})\dot{\bm{q}}
    \label{eq:v_m}
\end{equation}
\begin{equation}
    \bm{\omega}_j = \bm{\omega}_b + \bm{J}_{\bm{\omega}_j}(\bm{q})\dot{\bm{q}}
    \label{eq:w_m}
\end{equation}
where $\bm{J}_{\bm{v}_j}(\bm{q}) \in \mathbb{R}^{3 \times n}$ and $\bm{J}_{\bm{\omega}_j}(\bm{q}) \in \mathbb{R}^{3 \times n}$ are the linear and angular velocity Jacobians of the manipulator and $\dot{\bm{q}} \in \mathbb{R}^n $ are the arm  joints angular velocities.
Under the assumptions of Sec.~\ref{sec:assumptions} the system potential energy is zero and the Lagrangian function consists only of the term representing the system kinetic energy:
\begin{equation}
        \mathcal{K} \!\!=\!\! \frac{1}{2}\!\! 
        \left[m_b\bm{v}_b^T \bm{v}_b \!+\! 
              \bm{\omega}_b^T \bm{I}_b \bm{\omega}_b \!+\!\! 
              \sum_{j=1}^n m_j \bm{v}_j^T \bm{v}_j \!+\!\! 
              \sum_{j=1}^n \bm{\omega}_j^T \bm{I}_j \bm{\omega}_j \!
        \right]\!\!
    \label{eq:kineticEnergy}
\end{equation}
Substituting the relations~\eqref{eq:v_m} and~\eqref{eq:w_m} inside~\eqref{eq:kineticEnergy}, and grouping with respect to vector $ \dot{\bm{x}} = [\bm{v}_b^T, \bm{\omega}_b^T, \dot{\bm{q}}^T]^T \in \mathbb{R}^{6+n}$ allows to find a compact matrix formulation for the kinetic energy:}
{Under the assumptions above, \rev{the potential energy $\mathcal{V}$ of the system is zero and only the kinetic energy term $\mathcal{K}$ appears in the Lagrangian 
$\mathcal{L} = \mathcal{K}-\mathcal{V}$:}}
\begin{equation}
    \!
    \rev{\mathcal{L}} = \mathcal{K} = \frac{1}{2} \dot{\bm{x}}^T
\begin{bmatrix}
    \bm{M}_t & \bm{M}_{tr} & \bm{M}_{tm}\\
    \bm{M}_{tr}^T & \bm{M}_r & \bm{M}_{rm}\\
    \bm{M}_{tm}^T & \bm{M}_{rm}^T & \bm{M}_m\\
    \end{bmatrix}
    \dot{\bm{x}}
    = \frac{1}{2} \dot{\bm{x}}^T\bm{M}\dot{\bm{x}}
    \label{eq:kinEnergyOld}
\end{equation}
\LSv{where $\bm{M}=\bm{M}(\bm{q}_b,\bm{q}) \in \mathbb{R}^{(6+n)\times (6+n)}$, and specifically:}{\rev{where, $\dot{\bm{x}} = [\bm{v}_b^T, \bm{\omega}_b^T, \dot{\bm{q}}^T]^T \in \mathbb{R}^{6+n}$ and $\bm{M}=\bm{M}(\bm{q}_b,\bm{q}) \in \mathbb{R}^{(6+n)\times (6+n)}$: according to the notation above, $\bm{v}_b$ and $\bm{\omega}_b$ are linear and angular velocities of the BS w.r.t. the inertial frame, and $\dot{\bm{q}}$ velocity vector of the arm joints.} Specifically:}
\begin{subequations}
\small
    \begin{align}
   &\bm{M}_t = m_{tot}\bm{E} \in \mathbb{R}^{3 \times 3} \\
   &\bm{M}_{tr} = \sum_{j=1}^n [\bm{p}_{bj}]_\times m_j \in \mathbb{R}^{3 \times 3} \\ 
   &\bm{M}_r = \bm{I}_b + \sum_{j=1}^n \left( \bm{I}_j - m_j [\bm{p}_{bj}]_\times [\bm{p}_{bj}]_\times \right) \in \mathbb{R}^{3 \times 3} \\
   & \bm{M}_{tm} = \sum_{j=1}^n m_i \bm{J}_{\bm{v}_j} \in \mathbb{R}^{3 \times n}\\
   & \bm{M}_{rm} = \sum_{j=1}^n \left( m_j \bm{J}_{{\bm{\omega}}_j} + m_j [\bm{p}_{bj}]_\times \bm{J}_{\bm{v}_j} \right) \in \mathbb{R}^{3 \times n}\\
   & \bm{M}_{m} = \sum_{j=1}^n \left( \bm{J}_{\bm{\omega}_j}^T \bm{I}_j \bm{J}_{\bm{\omega}_j} + m_j \bm{J}_{\bm{v}_j}^T \bm{J}_{\bm{v}_j} \right) \in \mathbb{R}^{n \times n}
   \end{align}
\end{subequations}
with $m_{tot}$ being the total mass of space manipulator \LSv{}{and $\bm{J}_{\bm{v}_j}(\bm{q}) \in \mathbb{R}^{3 \times n}$ and $\bm{J}_{\bm{\omega}_j}(\bm{q}) \in \mathbb{R}^{3 \times n}$ the robotic arm linear and angular velocity Jacobians}. However, it is not possible to use $\bm{x}$ as generalized coordinate vector since the integral of the angular velocity does not have physical meaning. To overcome this issue, it is possible to consider a change of variable and to describe the BS angular velocity $\bm{\omega}_b$ in terms of quaternion derivative $\dot{\bm{q}}_b$, as:
%
\begin{equation}
    \bm{\omega}_b = 2 \begin{bmatrix} -\bm{\epsilon}_b \quad \eta_b \bm{E}+[\bm{\epsilon}_b]_\times \end{bmatrix} \dot{\bm{q}}_b = 2 \bm{G}(\bm{q}_b) \dot{\bm{q}}_b 
    \label{eq:quat_dot2w}
\end{equation}
where $\bm{G}(\bm{q}_b)\in \mathbb{R}^{3 \times 4}$.
Hence, using~\eqref{eq:quat_dot2w} we can rewrite $\dot{\bm{x}}$ as function of the BS  attitude quaternion derivative:
\begin{equation}
   \dot{\bm{x}} = 
   \begin{bmatrix}
   \bm{E} & \bm{0}& \bm{0}\\
   \bm{0} & 2 \bm{G}(\bm{q}_b) & \bm{0}\\
   \bm{0} & \bm{0}& \bm{E}\\
   \end{bmatrix} 
   \begin{bmatrix}
   \bm{v}_b\\
   \dot{\bm{q}}_b\\
   \dot{\bm{q}}
   \end{bmatrix} = \bm{H}(\bm{q}_b) \dot{\tilde{\bm{x}}}
   \label{eq:changeOfVar}
\end{equation}
where $\bm{H}(\bm{q}_b)\in \mathbb{R}^{(6+n) \times (7+n)}$ and  $\dot{\tilde{\bm{x}}} \in  \mathbb{R}^{7+n}$.
Substituting~\eqref{eq:changeOfVar} inside~\eqref{eq:kinEnergyOld} allows to obtain a description of the kinetic energy based on $\dot{\tilde{\bm{x}}} = [\bm{v}_b^T, \dot{\bm{q}}_b^T, \dot{\bm{q}}^T]^T \in  \mathbb{R}^{7+n}$, the integral of which is well defined from a physical point of view, and therefore it can be used as a generalized coordinates vector: $\tilde{\bm{x}} = [\bm{p}_b^T, \bm{q}_b^T, \bm{q}^T]^T \in  \mathbb{R}^{7+n}$. Clearly, such vector will make sense only if the values of $\bm{q}_b$ are chosen in $\mathbb{S}^3$.
The kinetic energy equation in this new formulation becomes:
\begin{equation}
    \mathcal{K} = \frac{1}{2} \dot{\tilde{\bm{x}}}^T \bm{H}^T(\bm{q}_b) \bm{M}(\bm{q}_b,\bm{q}) \bm{H}(\bm{q}_b) \dot{\tilde{\bm{x}}} =  \frac{1}{2} \dot{\tilde{\bm{x}}}^T \tilde{\bm{M}}(\tilde{\bm{x}}) \dot{\tilde{\bm{x}}}
    \label{eq:kinEnergy}
\end{equation}
where $\tilde{\bm{M}}(\bm{\tilde{x}}) \in \mathbb{R}^{(7+n)\times (7+n)}$. 
Applying the Lagrangian method on~\eqref{eq:kinEnergy} allows to find the system dynamical model:
\begin{equation}
    \tilde{\bm{M}}(\tilde{\bm{x}})\Ddot{\tilde{\bm{x}}} + \tilde{\bm{C}}(\dot{\tilde{\bm{x}}},\tilde{\bm{x}}) = \tilde{\bm{F}}
    \label{eq:dynSysTemp}
\end{equation}
where the vector $\Ddot{\tilde{\bm{x}}} = [\bm{a}_b^T, \Ddot{\bm{q}}_b^T, \Ddot{\bm{q}}^T]^T \in  \mathbb{R}^{7+n}$ stacks the BS linear and quaternion accelerations w.r.t. the $\mathcal{I}$ frame and the ones of arm joints; $\tilde{\bm{M}}(\tilde{\bm{x}})$ and $\tilde{\bm{C}}(\dot{\tilde{\bm{x}}},\tilde{\bm{x}}) \in \mathbb{R}^{(7+n)}$ are the inertia and Coriolis/Centrifugal matrices respectively and $\tilde{\bm{F}} \in \mathbb{R}^{7+n}$ is the generalized forces vector. 
Usually generalized forces and accelerations are expressed in terms of $\bm{F} = [\bm{f}^T_b,\bm{\tau}^T_b,\bm{\tau}^T]^T \in \mathbb{R}^{6+n}$ and $\Ddot{\bm{x}} = [\bm{a}_b^T, \dot{\bm{\omega}}_b^T, \Ddot{\bm{q}}^T]^T \in  \mathbb{R}^{6+n}$ respectively. 
The relation between $\tilde{\bm{F}}$ and $\bm{F}$ is:
\begin{equation}
    \tilde{\bm{F}} =  \bm{H}^T(\bm{q}_b) \bm{F}
    \label{eq:genForChange}
\end{equation}
and that between $\Ddot{\tilde{\bm{x}}}$ and $\Ddot{\bm{x}}$ is:
\begin{equation}
    \Ddot{\tilde{\bm{x}}} =   \begin{bmatrix}
    \bm{E} & \bm{0} & \bm{0}\\
    \bm{0} & 0.5\bm{G}(\bm{q}_b)^T & \bm{0}\\
    \bm{0} & \bm{0} & \bm{E} 
    \end{bmatrix}  \Ddot{\bm{x}} + \begin{bmatrix}
    \bm{0}\\
    \dot{\bm{G}}(\dot{\bm{q}_b})^T\bm{G}(\bm{q}_b)\dot{\bm{q}_b}\\
    \bm{0}
    \end{bmatrix}
    \label{eq:stateChange}
\end{equation}
Using~\eqref{eq:genForChange} and~\eqref{eq:stateChange} inside~\eqref{eq:dynSysTemp}, and exploiting the relation $\bm{G}(\bm{q}_b)\bm{G}^T(\bm{q}_b) = \bm{E}$, it follows:
\begin{equation}
    \bm{M}(\tilde{\bm{x}})\Ddot{\bm{x}} + \bm{C}(\dot{\bm{x}},\tilde{\bm{x}}) = \bm{F}
    \label{eq:dynSys}
\end{equation}
where $\bm{M}(\tilde{\bm{x}})$ is that of~\eqref{eq:kinEnergyOld} and $\bm{C}(\dot{\bm{x}},\tilde{\bm{x}})$ embeds the term $\bm{Q}(\dot{\bm{q}}_b,\bm{q}_b)$.
%
%
$\bm{F}$ is composed of one component associated with the actuated control inputs $\bm{u}_c \in \mathbb{R}^{6+n}$ and another related to unmodeled disturbances $\bm{d} \in \mathbb{R}^{6+n}$.

\section{Non singular terminal SMC}
The SMC rationale is to confine the closed loop dynamics to a subspace of the state space, called sliding surface, with interesting convergence properties. In the case of NTSMC, the selected subspace endows the system of finite time convergence capability and does not present any singularities.

\label{sec:contr}

\subsection{Sliding surface for translational and arm joints dynamics}
\label{sec:slidTrJ}

The following non singular terminal sliding surface is considered for controlling the translational dynamics of the BS and the one of the joints of the manipulator:
\begin{equation}
    \bm{s}_1 = \bm{\Gamma}_1^{-1} \dot{\bm{e}}^\frac{q}{p}+\bm{e}
    \label{eq:sliding1}
\end{equation}
where $\bm{e} = \begin{bmatrix} \bm{p}_b - \bm{p}_{b,d} \\ \bm{q} -\bm{q}_d  \end{bmatrix} \in \mathbb{R}^{3+n}$, $\dot{\bm{e}} = \begin{bmatrix} \bm{v}_b - \bm{v}_{b,d} \\ \dot{\bm{q}} - \dot{\bm{q}}_d  \end{bmatrix} \in \mathbb{R}^{3+n} $, 
$p$ and $q$ are positive odd integers that satisfy the relation $1<q/p<2$ and $\bm{\Gamma}_1 \in \mathbb{R}^{(3+n) \times (3+n)}$ is a diagonal positive definite matrix. To ease the notation the tracking errors are expressed w.r.t. the $\mathcal I$ and the reference quantities are marked by the additional subscript $d$ . 

\begin{theorem}
Assuming the manifold of the sliding surface $\mathcal{S}_1=\{ (\bm{e},\dot{\bm{e}} )\text{ s.t.  } \bm{s}_1(\bm{e},\dot{\bm{e}})=0 \}$ to be invariant, then the $i^{th}$ component of the error $e_i$ converges to zero in finite time.

Proof. \normalfont{The assumption on $\mathcal{S}_1$ invariance translates into the constraint $\bm{s}_1=0, \ \forall t > t_{r_1}$, where $t_{r_1}$ is the instant when the sliding surface is reached. Substituting this in~\eqref{eq:sliding1} allows finding the error dynamics when the sliding mode is enforced. The $i^{th}$ component of the error $\bm{e}$ evolves as:
\begin{equation}
    \dot{e}_i = - \Gamma_i^{\frac{p}{q}} e_i^{\frac{p}{q}}
    \label{eq:erridyn}
\end{equation}
Solving \eqref{eq:erridyn} leads to an explicit formulation of the $i^{th}$ component error convergence time:
\begin{equation}
    t_{s_1}^i = \frac{1}{\Gamma_i^{\frac{p}{q}}} \int^{0}_{e_i(t_{r_1})} \frac{d e_i}{e_i^{\frac{p}{q}}} = \frac{\lvert e_i(t_{r_1}) \lvert^{1-\frac{p}{q}}}{\Gamma_i^{\frac{p}{q}}(1-\frac{p}{q})}
\end{equation} 
which is clearly finite.\qed
}
\end{theorem}

The time derivative of $\bm{s}_1$, which will be used later on in the stability analysis of the controller, is:
%
%
\begin{equation}
    \!\dot{\bm{s}}_1 \!=\! \bm{\Gamma}^{-1} \frac{q}{p} \text{diag}\left(\dot{\bm{e}}^{\frac{q}{p}-1} \right)\Ddot{\bm{e}}\!+\!\dot{\bm{e}}
\quad \textrm{with} \quad 
\Ddot{\bm{e}}\!=\!\begin{bmatrix} \bm{a}_b \!-\! \bm{a}_{b,d} \\ \Ddot{\bm{q}} \!-\!\Ddot{\bm{q}} _d  \end{bmatrix}.
    \label{eq:sliding1dot}
\end{equation}

\subsection{Sliding surface for rotational dynamics}
\label{sec:slidRt}

The sliding surface used to control the rotation of the BS has a structure similar to \eqref{eq:sliding1}, with different error quantities:
\begin{equation}
    \bm{s}_2 = \bm{\Gamma}_2^{-1} (\bm{\omega}_{b,e})^\frac{q}{p}+\text{sgn}_+(\eta_{b,e})\bm{\epsilon}_{b,e}
    \label{eq:sliding2}
\end{equation}
where $\bm{\omega}_{b,e}=\bm{\omega}_{b}-\bm{\omega}_{b,d}  \in \mathbb{R}^3$, the function $\text{sgn}_+(\cdot)$ assumes the value $-1$ for the components of the input vector that are negative and $+1$ for the others, and $\bm{q}_{b,e} = [\eta_{b,e}, \bm{\epsilon}_{b,e}^T]^T = \bm{q}_b\otimes \bm{q}_{b,d}^{-1} \in \mathbb{S}^3$. 
In addition, $p$ and $q$ are the same positive odd integers considered for $\bm{s}_1$, and $\bm{\Gamma}_2 = \gamma_2\bm{E} \in \mathbb{R}^{3 \times 3}$. 
The term $\text{sgn}_+(\cdot)$ is added to prevent the unwinding phenomenon related to quaternions' dual coverage so that the shortest rotation is always selected~\cite{unwinding}.  

\begin{lemma}
\label{lemma:fintimeconLyap}
\cite{smNTSMC} Consider a system of the form $\dot{\bm{x}} = \bm{f}(\bm{x})$ with state $\bm{x} \in \mathbb{R}^n$ s.t. 
$\bm{f}(\bm{0}) = \bm{0}$ that has a unique solution in forwarding time for all initial conditions. If there exist a Lyapunov function $V(\bm{x})$ and two positive constants $\lambda$ and $\alpha \in (0,1)$ such that $\dot{V}(\bm{x}) \leq - \lambda V^\alpha(\bm{x})$ then $\bm{x}$ can be stabilized to the origin in finite time. Also, given the initial state $\bm{x}_0$, the settling time results to be:
\begin{equation}
    T \leq \frac{V(\bm{x}_0)^{1-\alpha}}{\lambda(1-\alpha)} 
\end{equation}
\end{lemma}

\begin{lemma}
    \label{lemma:ineq}
     \cite{disugSlidSurf} For any set of real numbers $x_i \in \mathbb{R}$ and positive constant $p \in (0,1]$ the following inequality holds:
    \begin{equation}
        \label{lemma:quat}
        \sum^n_{i=1} \vert x_i \vert^{1+p} \geq \left( \sum^n_{i=1} \vert x_i \vert^2 \right)^{\frac{1+p}{2}}
    \end{equation}
\end{lemma}

\begin{theorem}
Assuming the manifold of the sliding surface $\mathcal{S}_2=\{ (\bm{\omega}_{b,e},\bm{\epsilon}_{b,e})\text{ s.t.  } \bm{s}_2(\bm{\omega}_{b,e},\bm{\epsilon}_{b,e})=0 \}$ to be invariant, then the error quaternion converges to the identity quaternion in finite time.

Proof.  \normalfont{
The invariance of $\mathcal{S}_2$ translates into the constraint $\bm{s}_2=0, \ \forall t > t_{r_2}$, where $t_{r_2}$ is when the sliding surface is reached. 
The error convergence time can be found by employing the following Lyapunov function:
\begin{equation}
    \begin{split}
       V(\bm{\epsilon}_{b,e}) & = \bm{\epsilon}_{b,e}^T\bm{\epsilon}_{b,e}
    \end{split}
    \label{eq:lyaps2}
\end{equation}
Taking the time derivative of \eqref{eq:lyaps2} and using \eqref{eq:sliding2} together with the constraint $\bm{s}_2=0$, by Lemma~\ref{lemma:ineq} and the equality $\dot{\bm{\epsilon}}_{b,e} = 0.5(\eta_{b,e}\bm{\omega}_{b,e} -\left[\bm{\epsilon}_{b,e} \right]_\times\bm{\omega}_{b,e})$, we obtain:
\begin{equation}
    \begin{split}
        \dot{V}(\bm{\epsilon}_{b,e}) &= \eta_{b,e}\bm{\omega}_{b,e}^T\bm{\epsilon}_{b,e}\\ &= -\eta_{b,e}\text{sgn}_+(\eta_{b,e}) \gamma_2^\frac{p}{q} \left(\bm{\epsilon}_{b,e}^T\right)^{\frac{p}{q}} \bm{\epsilon}_{b,e}\\
        &\leq \gamma_2^\frac{p}{q} \text{min}(\lvert \eta_{b,e} \lvert) V(\bm{\epsilon}_{b,e})^\frac{1+\frac{p}{q}}{2}\\
    \end{split}
\end{equation}
Once the sliding mode is enforced, according to Lemma~\ref{lemma:fintimeconLyap}, the error quaternion converges to $(\pm 1, \bm{0})$ in finite time:
\begin{equation}
    t_{s_2} \leq \frac{V(\bm{\epsilon}_{b,e}(t_{r_2}))^{1-\frac{p}{q}}}{\gamma_2^\frac{p}{q} \text{min}(\lvert \eta_{b,e} \lvert)\left(1-\frac{p}{q}\right)}
\end{equation}
If $\text{min}(\lvert \eta_{b,e} \lvert)=0$, i.e. if $\norm{\bm{\epsilon}_{b,e}}^2=1$, using \eqref{eq:sliding2} it follows that $\dot{\bm{\epsilon}}_{b,e}>0$ and hence $\eta_{b,e}$ instantly moves away from such condition.
\qed
}
\end{theorem}

As before, the sliding surface derivative is computed as:
\begin{equation}
    \dot{\bm{s}}_2 = \bm{\Gamma}_2^{-1}\frac{q}{p} \text{diag}\left(\bm{\omega}_{b,e}^{\frac{q}{p}-1}\right)\dot{\bm{\omega}}_{b,e}+\text{sgn}_+(\eta_{b,e})\dot{\bm{\epsilon}}_{b,e}
    \label{eq:sliding2dot}
\end{equation}
It is to note that \eqref{eq:sliding2dot} is not defined for $\eta_{b,e}=0$: however, in this case, the controller proposed in next section generates a control input different than zero that moves the system.
\subsection{Control design}

Rearranging 
\eqref{eq:dynSys}, it is possible to describe the accelerations of the system in terms of control inputs and disturbances:
\begin{equation}
        \Ddot{\bm{x}} = -\bm{M}^{-1}(\tilde{\bm{x}})\bm{C}(\dot{\bm{x}},\tilde{\bm{x}}) + \bm{M}^{-1}(\tilde{\bm{x}}) (\bm{u}_c +\bm{d})
        \label{eq:dynSysAcc}
\end{equation}
where matrix $\bm{M}(\tilde{\bm{x}})$ is always invertible for physically consistent inertial parameters. Unfortunately, though, in a real world scenario matrices $\bm{M}(\tilde{\bm{x}})$ and $\bm{C}(\dot{\bm{x}},\tilde{\bm{x}})$ are uncertain. Indeed, due to the capture of an unknown target or unmodeled behaviors (e.g. wear, actuator dynamics, fuel sloshing, frictions), it is not possible to perfectly know the system's parameters. 
The issue can be addressed by splitting $\bm{M}(\tilde{\bm{x}})$ and $\bm{C}(\dot{\bm{x}},\tilde{\bm{x}})$ into modeled and unmodeled components, labeled with subscripts $0$ and subscript $\Delta$ respectively:
\begin{subequations}
    \begin{align}
    &\bm{M}(\tilde{\bm{x}}) = \bm{M}_0(\tilde{\bm{x}}) + \bm{M}_{\Delta}(\tilde{\bm{x}})\\
    &\bm{C}(\dot{\bm{x}},\tilde{\bm{x}}) = \bm{C}_0(\dot{\bm{x}},\tilde{\bm{x}}) + \bm{C}_\Delta(\dot{\bm{x}},\tilde{\bm{x}})
    \end{align}
    \label{eq:unmodeledSys}
\end{subequations}
\hspace{-8pt} Combining \eqref{eq:dynSysAcc}-\eqref{eq:unmodeledSys}, and introducing $\bm{\delta} \in \mathbb{R}^{6+n}$ to collect all external disturbances and unmodeled dynamics we get: 
\begin{equation}
        \Ddot{\bm{x}} = -\bm{M}_0^{-1}(\tilde{\bm{x}}) \bm{C}_0(\dot{\bm{x}},\tilde{\bm{x}}) + \bm{M}_0^{-1}(\tilde{\bm{x}}) {\bm{u}}_c +\bm{\delta} \\
        \label{eq:uncDynSysAcc}
\end{equation}

\begin{assumption}
\label{ass:boundedDis}
Each element $\delta_i$ of vector $\bm{\delta}$, which groups all uncertainties, can be strictly upper-bounded by a positive function of the velocity measurements \cite{jia2020continuous}.
\begin{equation}
    \vert {\delta}_i \vert < (\gamma_{1_i} + \gamma_{2_i} \norm{\dot{\bm{x}}}^2) <k_{\delta_i}(1+\norm{\dot{\bm{x}}}^2)
\end{equation}
where $\gamma_{1_i}$ and $ \gamma_{2_i} \in [1, \dots, 6+n]$ are positive constants and $k_{\delta_i} = max \{ \gamma_{1_i},\gamma_{2_i} \}  $.
\end{assumption}

To jointly control all the degrees of freedom of the space manipulator, the two sliding surfaces $\bm{s}_1$ and $\bm{s}_2$ 
are combined in a single vector $\bm{s}$, reordered to match the structure of the generalized acceleration vector $\Ddot{\bm{x}} = [\bm{a}_b^T, \dot{\bm{\omega}}_b^T, \Ddot{\bm{q}}^T]^T$:
%
\begin{equation}
    \bm{s} = 
    \begin{bmatrix}
    \bm{E}_{3 \times 3} & \bm{0} & \bm{0}\\
    \bm{0} & \bm{0} & \bm{E}_{3 \times 3}\\
    \bm{0} & \bm{E}_{n \times n} & \bm{0}\\
    \end{bmatrix}
    \begin{bmatrix}
    \bm{s}_1\\
    \bm{s}_2
    \end{bmatrix}
     = 
     \bm{P}\begin{bmatrix}
    \bm{s}_1\\
    \bm{s}_2
    \end{bmatrix}
\end{equation}
where $\bm{s} \in \mathbb{R}^{6 + n}$ is the sliding surface of the entire system and ${\bm{P} \in \mathbb{R}^{(6+n) \times (6+n)}}$. 
%
Moreover, the chattering phenomenon is 
here tackled through the boundary layer thickness method~\cite{unwinding}, and, 
to this aim, $\bm{s}$ is replaced with
\begin{equation}
    \Delta \bm{s} = \bm{s} - \Phi \text{ sat}\left(\frac{\bm{s}}{\Phi}\right)
    \label{eq:deltas}
\end{equation}
where $\Phi$ is a positive constant and the sat($\cdot$) function bounds each component of its input vector between $\pm 1$. 
The result is that $i^{th}$ component of $\Delta \bm{s}$ goes to zero if the corresponding term of $\bm{s}$ is lower than $\Phi$. 
In addition, the switching term of the controller, which confers robustness against disturbances and unmodelled aspects, is based on the sat($\cdot$) function defined above and not on the classical sgn($\cdot$) one. 
Clearly, from \eqref{eq:deltas} it is easy to see that if $\lvert s_i \lvert \ \geq \Phi \ \forall i \in [1, \dots, 6+n]$ then $\dot{\Delta \bm{s}} = \dot{\bm{s}}$ and $\Phi$sat($\bm{s}/{\Phi}$) =  sgn($\Delta \bm{s}$). 
The basic idea underneath this technique is to consider the sliding mode to be enforced in a hypercylinder of radius $\Phi$ from the actual sliding surface and to modulate the switching term inside this hyperspace proportionally to the distance from the surface itself.
By doing so, the controller becomes less aggressive 
when close to the sliding surface preventing the high frequency control switching typical of the chattering problem. 
In summary, the proposed controller has the  structure:
\begin{equation}
    \bm{u}_c = \bm{u}_1 +\bm{u}_2 + \bm{u}_3
    \label{eq:control}
\vspace{-3pt}
\end{equation}
with:
\begin{subequations}
\small
    \begin{align}
    & \!\!\!\!\!\bm{u}_1 \!=\! \bm{M}_0(\tilde{\bm{x}}) \begin{bmatrix}
    \bm{a}_{b,d}\\
    \dot{\bm{\omega}}_{b,d}\\
    \Ddot{\bm{q}}_d
    \end{bmatrix} + \bm{C}(\tilde{\bm{x}},\dot{\bm{x}})\\ 
    & \label{eq:u2} \!\!\!\!\!\bm{u}_2 \!=\! - \frac{p}{q}\bm{M}_0(\tilde{\bm{x}}) \bm{P}  \begin{bmatrix}
    \bm{\Gamma}_1 \dot{\bm{e}}^{2-\frac{q}{p}}\\
    \text{sgn}_+(\eta_{b,e}) \bm{\Gamma}_2 \ \text{diag}\left(\frac{\bm{\omega}_{b,e}^{\frac{q}{p}-1}}{\bm{\omega}_{b,e}^{\frac{2q}{p}-2}} \right) \dot{\bm{\epsilon}}_{b,e}\\
    \end{bmatrix}\\
    \begin{split}
        & \!\!\!\!\!\bm{u}_3 \!=\! - \bm{M}_0(\tilde{\bm{x}}) \left( \hat{\bm{K}}_{\bm{\delta}} \ \Phi \ \text{sat} \left(\bm{s}/{\Phi}\right) (1+\norm{\dot{\bm{x}}}^2) \right. \\
        & \!\!\!\!\!\!\qquad \ +  \left. \frac{\bm{\xi}}{\norm{\bm{\xi}}^2} \bm{K}_1 \norm{\Delta \bm{s}}^{2\frac{p_1}{q_1}} + \bm{K}_2 \ \Delta \bm{s}^{\frac{p_2}{q_2}} \right)
    \end{split}
    \end{align}
    \label{eq:controlInput}
\end{subequations}
where 
\begin{equation*}
    \bm{\xi}= \text{diag}\left( \bm{P} \begin{bmatrix}
    \bm{\Gamma}_1^{-1} \dot{\bm{e}}^{\frac{q}{p}-1}\\
    \bm{\Gamma}_2^{-1}  \bm{\omega}_{b,e}^{\frac{q}{p}-1}
    \end{bmatrix} \right) \Delta \bm{s}
\end{equation*}
and $p_1$ and $q_1$ are positive integers whose ratio is $1<\frac{q_1}{p_1}<2$, $p_2$ and $q_2$ are positive odd integers that satisfy $p_2<q_2$, $\bm{K}_1 \in \mathbb{R}^{(3+n) \times (3+n)}$ and $\bm{K}_2 \in \mathbb{R}^{3 \times 3}$ are positive definite diagonal matrices. 
The ratio inside the diagonal operator in \eqref{eq:u2} is performed component-wise, and the elements of the denominator are also lower-bounded at $\Phi$ in order to prevent excessive control efforts.
The last element to define is the adaptive matrix $\hat{\bm{K}}_{\bm{\delta}}\in \mathbb{R}^{(6+n) \times (6+n)}$, namely the estimate of the diagonal matrix $\bm{K}_{\bm{\delta}}$ whose $i^{th}$ element corresponds to $k_{\delta_i}$ defined above. 
The adaptive law for $\hat{\bm{K}}_{\bm{\delta}}$ is:
\begin{equation}   \dot{\hat{\bm{K}}}_{\bm{\delta}} \!=\! \phi \ \text{diag}\left(\!\bm{P}\!\begin{bmatrix}
    \bm{\Gamma}_1^{-1} \dot{\bm{e}}^{\frac{q}{p}-1}\\
    \bm{\Gamma}_2^{-1}  \bm{\omega}_{b,e}^{\frac{q}{p}-1}
    \end{bmatrix}\!\right) \text{diag}(\vert \Delta \bm{s} \vert)(1+\norm{\dot{\bm{x}}}^2)
    \label{eq:adaptTerm}
\vspace{-3pt}
\end{equation}
where $\phi$ is a positive constant. 
%

The first controller term $\bm{u}_1$ consists in a feed-forward action; the second, $\bm{u}_2$, is required to create the sliding surface $\Delta \bm{s}$ \rev{and the $\text{sgn}_+(\cdot)$ discontinuity compensates that of \eqref{eq:sliding2} to provide closed-loop smoothness~\cite{unwinding}}; the last term $\bm{u}_3$ is used to boost the convergence to the sliding surface and to deal with uncertainties.  


\begin{theorem}
Consider the space manipulator system  model~\eqref{eq:dynSys}. 
Under Assumption \ref{ass:boundedDis}, the controller \eqref{eq:control} is able to guarantee finite time convergence of the tracking error $\bm{e}$ to zero and orientation error quaternion $\bm{q}_{b,e}$ to the identity one. 
In addition, also the estimation error $\tilde{\bm{K}_{\bm{\delta}}} =  \bm{K}_{\bm{\delta}} - \hat{\bm{K}}_{\bm{\delta}}$ of the adaptive terms converges asymptotically to zero. \vspace{3mm}

\begin{figure*}[b!]
    \hrule
    \small
    \begin{align*}
        & \dot{V} = \frac{q}{p} \bm{\xi}^T \left(-\bm{M}_0^{-1}(\tilde{\bm{x}}) \bm{C}_0(\dot{\bm{x}},\tilde{\bm{x}}) + \bm{M}_0^{-1}(\tilde{\bm{x}}) {\bm{u}}_c +\bm{\delta} - \begin{bmatrix}
            \bm{a}_{b,d} \\ 
            \dot{\bm{\omega}}_{b,d}\\
            \Ddot{\bm{q}} _d
            \end{bmatrix}\right) + \Delta \bm{s}^T \begin{bmatrix} 
            \bm{v}_b - \bm{v}_{b,d} \\ 
            \text{sgn}_+(\eta_{b,e})\dot{\bm{\epsilon}}_{b,e}\\
            \dot{\bm{q}} -\dot{\bm{q}}_d
            \end{bmatrix}   
            - \frac{q}{p} \sum^n_{i=1} \xi_i \tilde{k}_{\delta_i} \text{ sgn}(\Delta \bm{s}_i) (1+\norm{\dot{\bm{x}}}^2)
        \\
        &\quad= \frac{q}{p} \bm{\xi}^T  \bm{\delta} - \frac{q}{p} \bm{\xi}^T \hat{\bm{K}}_{\bm{\delta}} \text{sgn} \left( \Delta \bm{s}\right)(1+\norm{\dot{\bm{x}}}^2) - \underbrace{ 
            \bm{K}_1 \norm{\Delta \bm{s}}^{2\frac{p_1}{q_1}}}_{\geq 0} - \underbrace{ \frac{q}{p} \bm{\xi}^T \bm{K}_2 \ \Delta \bm{s}^{\frac{p_2}{q_2}}}_{\geq 0}
            - \frac{q}{p} \sum^n_{i=1} \xi_i \tilde{k}_{\delta_i} \text{ sgn}(\Delta \bm{s}_i)(1+\norm{\dot{\bm{x}}}^2)  
        \\
        & \quad \leq \frac{q}{p} \bm{\xi}^T  \bm{\delta} - \left( \frac{q}{p} \bm{\xi}^T \hat{\bm{K}}_{\bm{\delta}} \text{sgn} \left( \Delta \bm{s}\right) + \frac{q}{p} \sum^n_{i=1} \xi_i \tilde{k}_{\delta_i} \text{ sgn}(\Delta \bm{s}_i) \right)(1+\norm{\dot{\bm{x}}}^2) \leq - \underbrace{\frac{q}{p} \sum^n_{i=1} \xi_i \text{ sgn}(\Delta \bm{s}_i)}_{\geq 0} \underbrace{ \left((\hat{k}_{\delta_i} + \tilde{k}_{\delta_i})(1+\norm{\dot{\bm{x}}}^2) - \vert \delta_i \vert \right)}_{> 0} \leq 0
    \end{align*}
    \vspace{-10mm}
\end{figure*}

Proof.
\normalfont{
Firstly, the asymptotic convergence of the sliding surface $\Delta \bm{s}$ is proven by evaluating the Lyapunov function:
\begin{equation}
V = \frac{\Delta\bm{s}^T\Delta\bm{s}}{2} + \frac{q}{2p} \phi^{-1} \sum^n_{i=1} \bm{\gamma}_i^{-1} \tilde{k}_{\delta_i}^2
\label{eq:lyapCont}
\end{equation}
where $\tilde{k}_{\delta_i}$ and $\bm{\gamma}_i^{-1}$ are the $i^{th}$ diagonal element of matrices $\tilde{\bm{K}_{\bm{\delta}}}$ and $\bm{P} \text{ diag} ( \bm{\Gamma}_1^{-1} \bm{\Gamma}_2^{-1})$ respectively. Under the assumption that $\bm{K}_{\bm{\delta}}$ does not change with time or it changes slowly with respect to the system dynamic, it holds that $\dot{\tilde{k}}_{\delta_i} = -\dot{\hat{k}}_{\delta_i}$. Taking the derivative of \eqref{eq:lyapCont}, we obtain:
\begin{equation}
    \dot{V} = \Delta \bm{s}^T \dot{\Delta \bm{s}}^T - \frac{q}{p} \phi^{-1} \sum^n_{i=1} \bm{\gamma}_i^{-1} \tilde{k}_{\delta_i} \dot{\hat{k}}_{\delta_i} 
    \label{eq:lyapContDot}
\end{equation}

Substituting \eqref{eq:sliding1dot}, \eqref{eq:sliding2dot}, and~\eqref{eq:adaptTerm} inside~\eqref{eq:lyapContDot} leads to:
\begin{equation}
    \begin{split}
        \!\dot{V} &= \frac{q}{p} \bm{\xi}^T\!\begin{bmatrix}
        \bm{a}_b - \bm{a}_{b,d} \\ 
        \dot{\bm{\omega}}_{b}-\dot{\bm{\omega}}_{b,d}\\
        \Ddot{\bm{q}} -\Ddot{\bm{q}} _d
        \end{bmatrix} \!+\! \Delta \bm{s}^T \!\begin{bmatrix}
        \bm{v}_b - \bm{v}_{b,d} \\ 
        \text{sgn}_+(\eta_{b,e})\dot{\bm{\epsilon}}_{b,e}\\
        \dot{\bm{q}} -\dot{\bm{q}}_d
        \end{bmatrix}\! \\
        &- \frac{q}{p} \sum^n_{i=1} \xi_i \tilde{k}_{\delta_i} \text{ sgn}(\Delta \bm{s}_i)(1+\norm{\dot{\bm{x}}}^2) 
    \label{eq:dim1}
    \end{split}
\end{equation}
where $\xi_i$ and $\Delta \bm{s}_i$ are the $i^{th}$ element of vectors $\bm{\xi}$ and $\Delta \bm{s}$ respectively. 
Using the dynamical model of the system \eqref{eq:uncDynSysAcc} and the controller \eqref{eq:control} inside \eqref{eq:dim1} yields to:
\begin{equation}
            \dot{V}  \leq - \underbrace{\frac{q}{p} \sum^n_{i=1} \xi_i \text{sgn}(\Delta \bm{s}_i)}_{\geq 0} \underbrace{ \left(k_{\delta_i}(1+\norm{\dot{\bm{x}}}^2) - \vert \delta_i \vert \right)}_{ > 0} \leq 0
            \label{eq:lyapDot1}
\vspace{-6pt}
\end{equation}
where in the last row Assumption~\ref{ass:boundedDis} and $k_{\delta_i} = \hat{k}_{\delta_i} + \tilde{k}_{\delta_i}$ are used and the inequality holds strict if  $\Delta \bm{s}_i$ is different than zero.
The main calculations necessary to derive the inequality~\eqref{eq:lyapDot1} are reported at the bottom of this page. 
The second step is to discuss the finite time convergence of $\Delta \bm{s}$: this is done by evaluating the following Lyapunov function: 
\begin{equation}
    V = \Delta \bm{s}^T \Delta \bm{s}/2
    \label{eq:lyapCont2}
\end{equation}
Following a procedure similar to the one just used for the Lyapunov function~\eqref{eq:lyapCont} it is easy to prove that:
\begin{equation}
    \begin{split}
        \dot{V} &\leq  -\frac{q}{p} \bm{K}_1 \norm{\Delta \bm{s}}^{2\frac{p_1}{q_1}} \\ 
        &-\frac{q}{p} \sum^n_{i=1} \xi_i \text{ sgn}(\Delta \bm{s}_i) \left(\hat{k}_{\delta_i} (1+\norm{\dot{\bm{x}}}^2) - \vert \delta_i \vert \right)
        \label{eq:lyapCont2dot}
    \end{split}
\end{equation}
if $\hat{k}_i (1+\norm{\dot{\bm{x}}}^2) \geq \delta_i \ \forall i \in [1,\dots,6+n]$ then it holds:
\begin{equation}
    \dot{V} \leq - \frac{q}{p} \bm{K}_1 2^{\frac{p_1}{q_1}} \left[ \left( \frac{\norm{\Delta \bm{s}}}{\sqrt{2}} \right)^2 \right]^\frac{p_1}{q_1} = - \frac{q}{p} \bm{K}_1 2^{\frac{p_1}{q_1}} V^\frac{p_1}{q_1}
    \label{eq:lyapCont2dot2}
\end{equation}
Now, applying Lemma~\ref{lemma:fintimeconLyap}, it is possible to conclude that the time $t_r$ needed to reach the sliding surface $\Delta \bm{s} = 0$ is:
\begin{equation}
    t_r = \frac{q_1}{ 2^{\frac{p_1}{q_1}} (q_1-p_1)\bm{K}_1 \frac{q}{p}} \left(\frac{\norm{\Delta \bm{s}(0)}}{\sqrt{2}}\right)^{\frac{q_1-p_1}{q_1}}
\end{equation}
This fact alongside the finite time convergence of the errors when the sliding mode is enforced concludes the proof. 
\qed
}
\end{theorem}


\LSv{
\section{Simulations and discussion}
\label{sec:sim}
Multiple numerical simulations are proposed to evaluate the performance of the control solution presented in this work. 
For comparison, the NTSMC designed here is compared with that proposed in \cite{smNTSMC} and a joint space Proportional Derivative (PD) controller based on feedback linearization.

The characteristics of the space manipulator considered in the simulations are defined according to those of previously accomplished space missions.\LSv{
The BS consists of a boxed shaped bus where two solar panels are connected and whose characteristics are reported in Tab~\ref{tab:BScha}. 
Moreover, the actuation system of the BS is supposed to be provided by the following set of devices:
\begin{itemize}
    \item  A group of four RWs, three of which are aligned with the three principal axes of inertia of the BS and the forth is positioned so that it is able to act equally along the three principal axes of inertia. 
    In addition, the maximum torque that can be generated by each RW is set to be $0.5 \ [Nm]$.
    \item A reaction control system made up of twenty-four thrusters able to produce independent torques and forces w.r.t. the three principal axes of inertia and positioned in a redundant configuration. 
    The nominal thrust and minimum impulse bit of each thruster is considered to be $10 \ [N]$ and $0.02 \ [Ns]$ respectively, their valves being driven by a PWPF (Pulse-Width Pulse-Frequency) modulator. 
\end{itemize}
The structure considered of the robotic arm
consists of a redundant seven degree of freedom configuration made up of only revolute joints, whose Denavit-Hartenberg parameters are given in Tab.~\ref{tab:DH}. 
The arm links are supposed to be realized with hollow aluminium cylinders whose thickness and external radius are $13.5 \ [mm]$ and $63.5 \ [mm]$ respectively. 
Both the DC motors of the arm and the EE are modelled as $0.5 \ [kg]$ point masses, and the maximum torque that each motor can produce is assumed to be $10 \ [Nm]$. }{The BS consists of a boxed shaped bus where two solar panels are connected and whose characteristics are reported in Tab~\ref{tab:BScha}. The actuation system of the BS is supposed to be made up by a group of four reaction wheels and a reaction control system. The structure considered of the robotic arm consists of a redundant seven degree of freedom configuration made up of only revolute joints, whose Denavit-Hartenberg parameters are given in Tab.~\ref{tab:DH}. The arm links are supposed to be realized with hollow aluminium cylinders and the DC motors of the arm and the EE are modelled as $0.5 \ [kg]$ point masses. Moreover, the main characteristics of the system actuators are modelled.}

\begin{table}[t]
    \centering
    \caption{BS characteristics}
    \begin{tabular}{|c|c|c|}
        \hline
        Mass & Dimensions & Inertias (COM)\\
        \hline
        \multicolumn{1}{|c|}{\multirow{4}{*}{$m_b = 1900 \ kg$}} & \multicolumn{1}{|c|}{\multirow{4}{*}{\shortstack{Depth $= 2.1 \ m$\\
         Length $= 2.3 \ m$ \\ Height $= 3.1 \ m$ }}} & $J_{xx} = 13500 \ kg m^2$\\
        & & $J_{yy} = 2000 \ kg m^2$\\ 
        & & $J_{yy} = 2000 \ kg m^2$\\ 
        & & $J_{ij} = 0 \ kg m^2$\\
        \hline
        
    \end{tabular}
    
    \label{tab:BScha}
\end{table}

\begin{table}[t]
    \centering
    \caption{Arm Denavit-Hartenberg parameters}
    \begin{tabular}{|c|c|c|c|c|c|c|c|}
    \hline
     & $l_1$ & $l_2$ & $l_3$ & $l_4$ & $l_5$ & $l_6$ & $l_7$ \\
    \hline
    $a_i \ [m]$ & 0 & 0 & 0 & 0 & 0 & 0 & 0 \\
    \hline
    $\alpha_i \ [rad]$ & $\pi/2$ & $-\pi/2$& $\pi/2$& $-\pi/2$& $\pi/2$
    &$-\pi/2$ & 0 \\
    \hline
    $d_i \ [m]$ & 0.3 & 0.16 & 1.15 & -0.16 & 1.15 & -0.16 & 0.4 \\
    \hline
    \end{tabular}
    \label{tab:DH}
\end{table}

The trajectory employed in all the simulations consists of two diagonal movements: one for BS and a following one for the EE. In this way all the degrees of freedom of the system are excited along the outlined movement path. In addition, the trajectory is chosen in such a way to have a good margin w.r.t. the saturation of the actuators in nominal condition.

In the following, three scenarios are analyzed with the purpose of showing the robustness of the solution presented in this work its increased fuel efficiency during attitude corrections. For evaluating the energy consumption the same metric used in~\cite{smNTSMC} is employed, namely  $E_{sim} = \int^{t_{sim}}_0 \bm{u}_{c,act}^T \bm{u}_{c,act} \ dt$ where $t_{sim}$ is the time span of simulation and $\bm{u}_{c,act}$ is the control action generated by the actuators. Even if, such index is not formally correct, since quantities with different unit of measure are summed up, it allows to compactly describes the control effort.

For a fair comparison, the parameters of the three controllers (reported in Tab.~\ref{tab:contParam}) are tuned so that all solutions perform and consume similarly in ideal conditions namely, if $\bm{u}_{c,act} = \bm{u}_{c}$ and no disturbances or uncertainties are present.

\begin{table}[t]
    \centering
    \caption{Parameters of the compared controllers}
    \begin{tabular}{|c|c|}
    \hline
    Controllers & Parameters\\
    \hline
    \multicolumn{1}{|c|}{\multirow{5}{*}{prop.  NTSMC}}    &  \multicolumn{1}{|c|}{\multirow{5}{*}{\shortstack{$p = 9$, $q=11$, $p_1=5$, $q_1=9$, $p_2=7$, $p_2=9$,\\ $\bm{\Gamma}_1 = 0.1  \bm{E}$, $\bm{\Gamma}_2 = 0.1 \bm{E}$, $\bm{K}_1= 10^{-2} \bm{E}$,\\ $\bm{K}_2= \text{diag}(0.1 \ \bm{1}_{1 \times 8},\ 0.2 \ \bm{1}_{1 \times 2}$,  $0.6 \ \bm{1}_{1 \times 4})$ \\ $\Phi= 10^{-3}$,  $\hat{\bm{K}}_{\bm{\delta}}(0) = 10^{-4}$, $\phi = 10^{-3}$ }}}\\ 
    & \\
    & \\
    & \\
    & \\
    \hline
    \multicolumn{1}{|c|}{\multirow{5}{*}{NTSMC \cite{smNTSMC}}}    &  \multicolumn{1}{|c|}{\multirow{5}{*}{\shortstack{$p = 9$, $q=11$, $p_2=5$, $q_2=9$, $p_3=7$, $p_3=9$,\\ $\bm{\beta}_1 = 0.1  \bm{E}$,  $k_0= 10^{-4}$, $k_1= 10^{-3}$,\\ $\bm{K}_2= \text{diag}(0.1 \ \bm{1}_{1 \times 8},\ 0.2 \ \bm{1}_{1 \times 2}$,  $0.6 \ \bm{1}_{1 \times 4})$\\ $\theta = 10^{-3}$, $c_0= 10^{-3}$}}}\\ 
    & \\
    & \\
    & \\
    & \\
    \hline
    \multicolumn{1}{|c|}{\multirow{3}{*}{PD}}    &  \multicolumn{1}{|c|}{\multirow{3}{*}{\shortstack{$\bm{K}_{\bm{p}_b} = 0.5  \bm{E}$, $\bm{K}_{\bm{v}_b} = 0.25  \bm{E}$, $\bm{K}_{\bm{q}_{\bm{\epsilon}_{b}}} = 0.5  \bm{E}$,\\ $\bm{K}_{\bm{\omega}_b} = 0.25  \bm{E}$, $\bm{K}_{\bm{q}} = 1  \bm{E}$, $\bm{K}_{\dot{\bm{q}}} = 0.5  \bm{E}$ }}}\\ 
    & \\
    & \\
    \hline
    \end{tabular}
    \label{tab:contParam}
\end{table}

\begin{figure}[t]
\centering
\includegraphics[trim = 13 0 30 0, clip,width=0.45\textwidth]{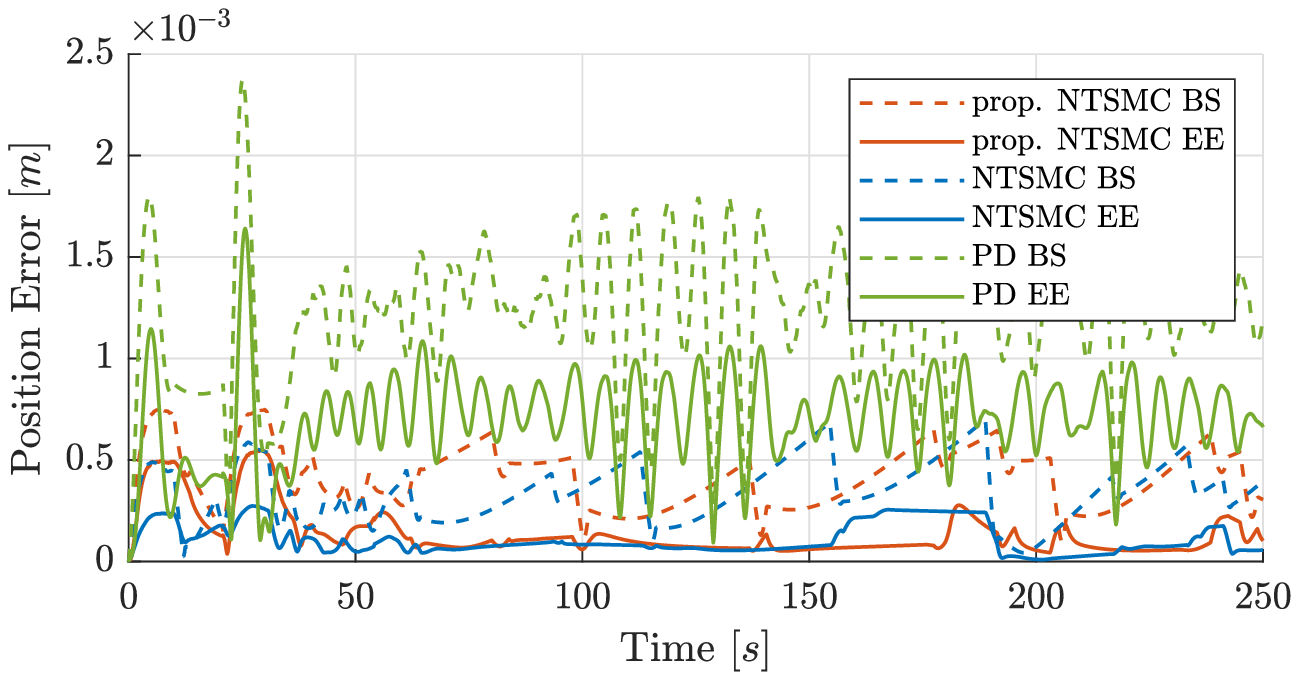}\\
\centering
\includegraphics[trim = 13 0 30 0, clip,width=0.45\textwidth]{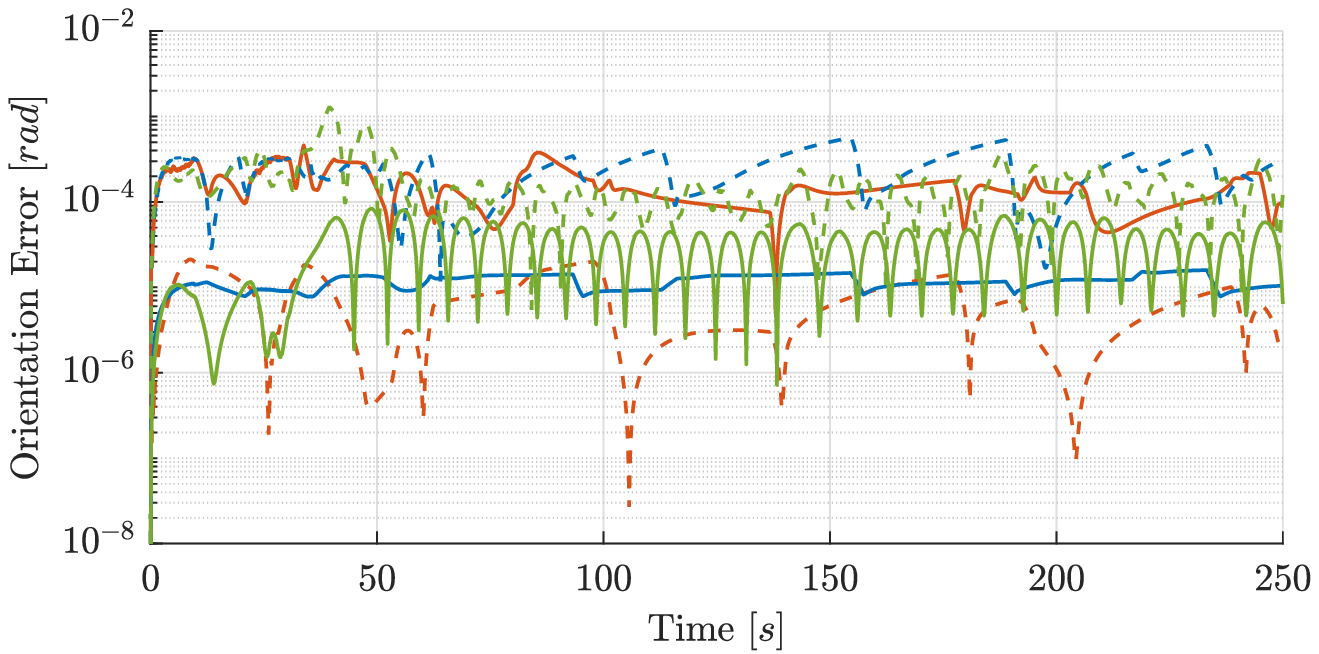}
\caption{ PD and NTSMC controllers. EE and BS tracking error norms.}
\label{fig:S1}
\end{figure}

Interestingly, when the realistic models of the actuators are included in simulation, even if all the controllers are able to achieve satisfying tracking performances in terms of position and orientation error norms (see Fig.~\ref{fig:S1}), it turns out that the two SMC solutions are more energy saving w.r.t. the PD one, as reported in Tab.~\ref{tab:S1_fuel}. 
\begin{table}[t]                                 
\centering     
\caption{Energy consumption}
\begin{tabular}{|c|c|c|c|}                        
\hline                                            
 & prop. NTSMC & NTSMC~\cite{smNTSMC} & PD \\                    
\hline                                            
$E_{sim}$ & 7233.81 & 7301.75 & 9366.98 \\          
\hline                                            
\end{tabular}                                     
\label{tab:S1_fuel}                            
\end{table}  
\begin{table}[t]                                                 
\centering
\caption{Energy consumption for attitude correction}
\begin{tabular}{|c|c|c|}                                        
\hline                                                        
& prop. NTSMC & NTSMC~\cite{smNTSMC} \\                                        
\hline                                                        
$E_{sim}$ & 105658.16 & 295103.52 \\                                       
\hline                                                        
\end{tabular}                                                 
\label{table:FuelCons2}                                        
\end{table} 
Therefore, the focus is then placed on the fuel efficiency and robustness features of the SMC control solutions.  
One main contribution of this work is to consider a sliding surface based on quaternions for controlling the attitude of the BS. Contrary to the solution proposed in \cite{smNTSMC} that is based on Euler Angles, the controller~\eqref{eq:controlInput} captures the topology of $\mathbb{S}^3$ and, as a consequence, promises to better handle attitude displacements of the BS, especially in the case of large ones. To verify this a scenario where the BS is rotated of $[-\frac{5}{4} \pi,0,-\frac{\pi}{4}]_{xyz}\ [rad]$ w.r.t. the reference trajectory is studied. From the plots in Fig.~\ref{fig:S1}, it is clear that the proposed solution is superior from a consumption point of view. Indeed, even if the convergence time of the attitude error norm is almost the same, the integral of torque norm required for the attitude correction is much lower when using the quaternion based solution and this results in a better fuel efficiency. Such an improved behaviour can be linked to the fact that controller designed in this work better captures the topology of $\mathbb{S}^3$. Indeed, looking at the vector part of the quaternion error, which is shown in Fig.~\ref{fig:S2}, it can be noticed that in the case of the proposed controller such error practically maintains its direction and scales its magnitude only. This means that the rotation axis is kept almost the same during error regulation, moreover the regulation path followed during attitude correction is close to the shortest one that represented in yellow.
\begin{figure*}

\begin{minipage}{0.63\linewidth}
    \begin{minipage}{0.5\linewidth}
    \centering
    \includegraphics[trim = 0 0 0 0, clip,width=1\textwidth]{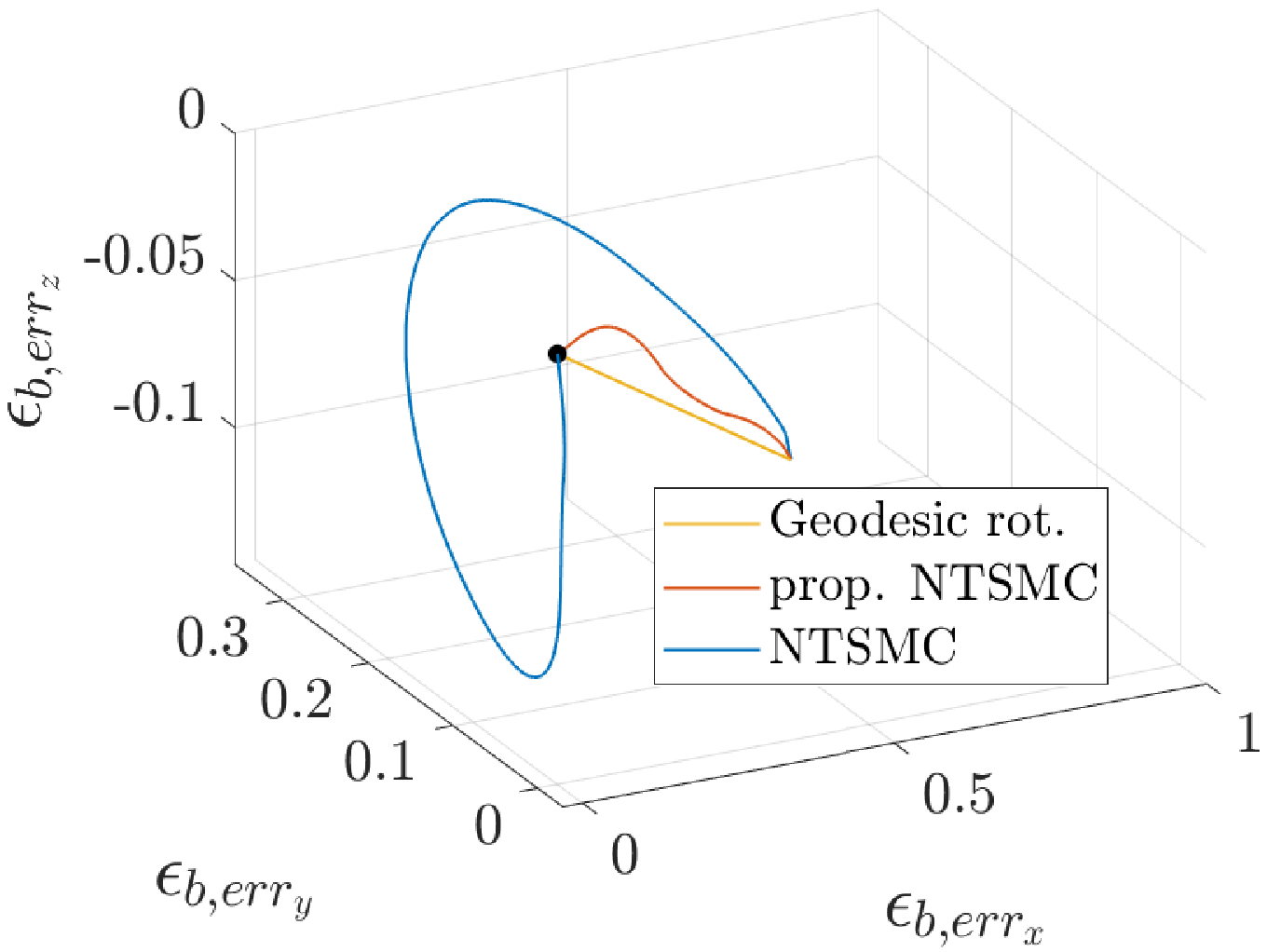}
    \end{minipage}
    \hfill
    \begin{minipage}{0.5\linewidth}
    \centering
    \includegraphics[trim = 10 0 45 0, clip,width=1\textwidth]{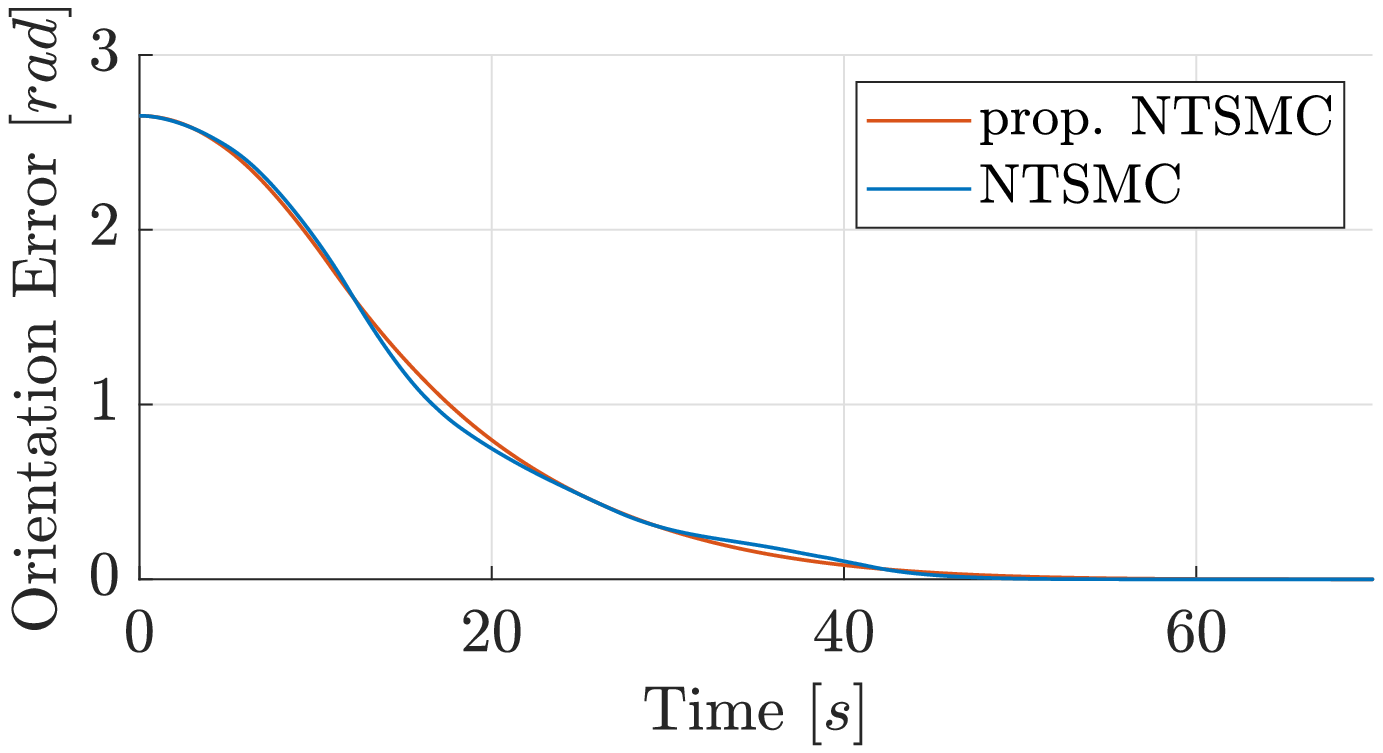}
    \includegraphics[trim = 10 0 45 0, clip,width=1\textwidth]{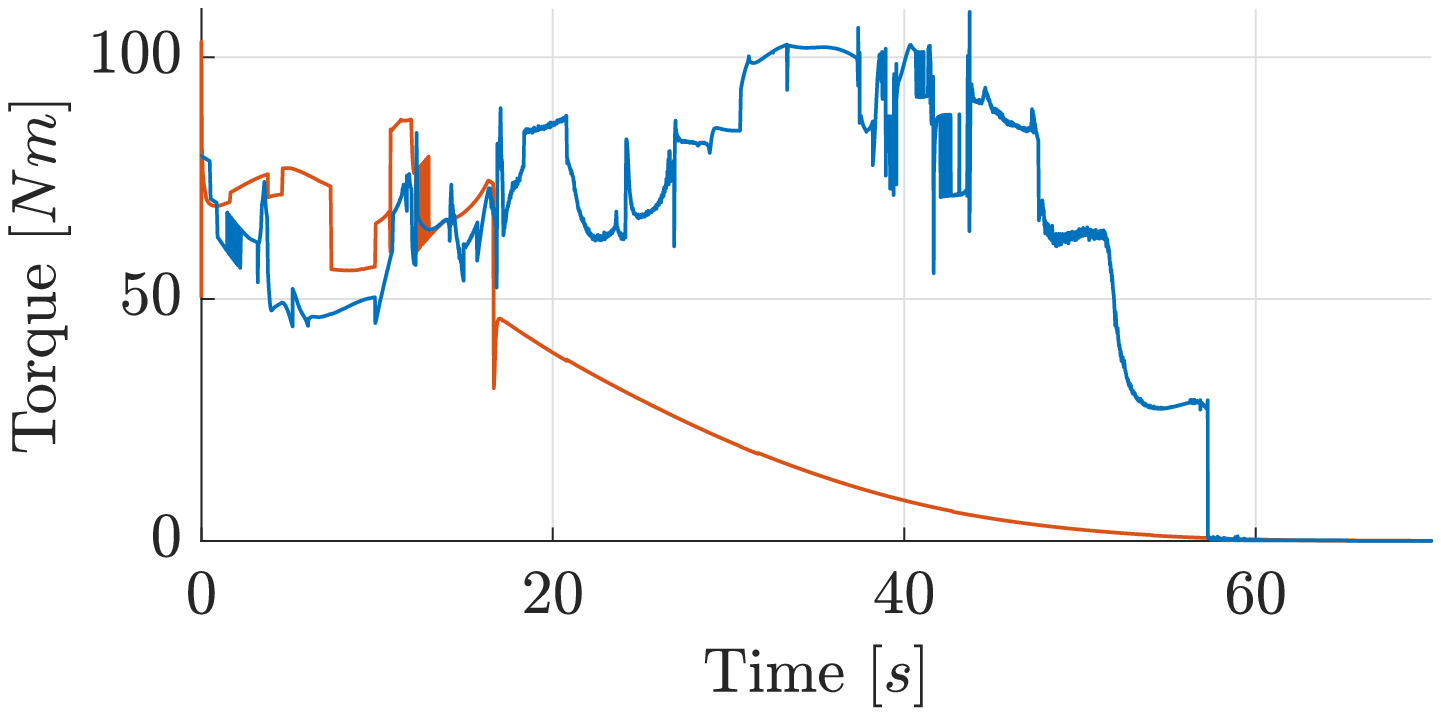}
    \end{minipage}
    \captionof{figure}{NTSMC controllers. On the left: evolution of the vector part of the error quaternion. Given the axis-angle parametization of the orientation error, $\bm{\epsilon}_{b,e}$ can be interpreted as the path of the rotation axis scaled by half the sine of the rotation angle. For this reasons it gives a hint on the rotation direction and its magnitude. For reference the geodesic rotation, i.e. the shortest one, is shown in yellow. On the right: BS attitude error and control torque norms.}
    \label{fig:S2}
\end{minipage}
\hfill
\begin{minipage}{0.34\linewidth}
\centering
\includegraphics[trim = 10 0 25 25, clip,width=1\textwidth]{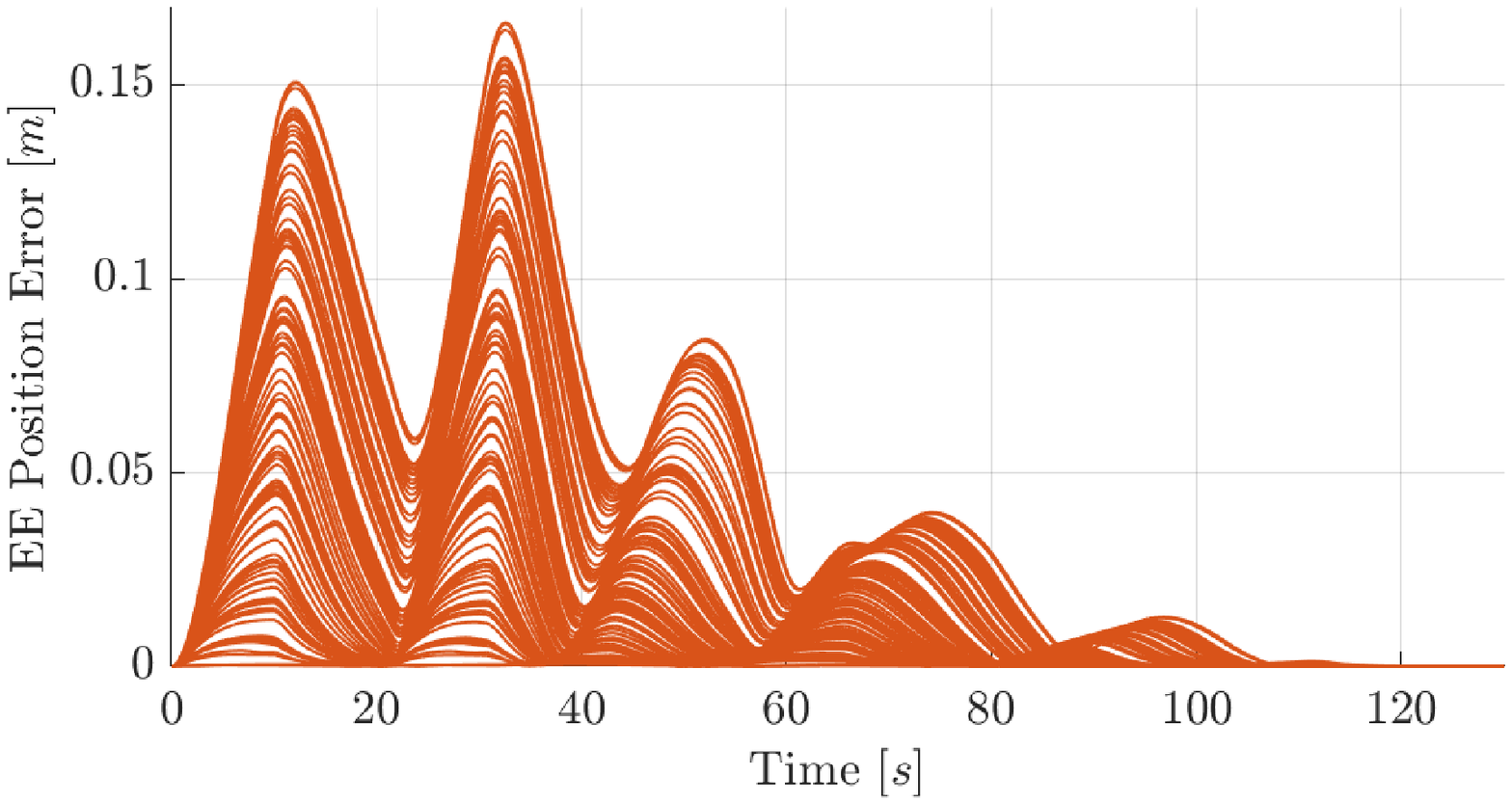}
\includegraphics[trim = 10 0 25 10, clip,width=1\textwidth]{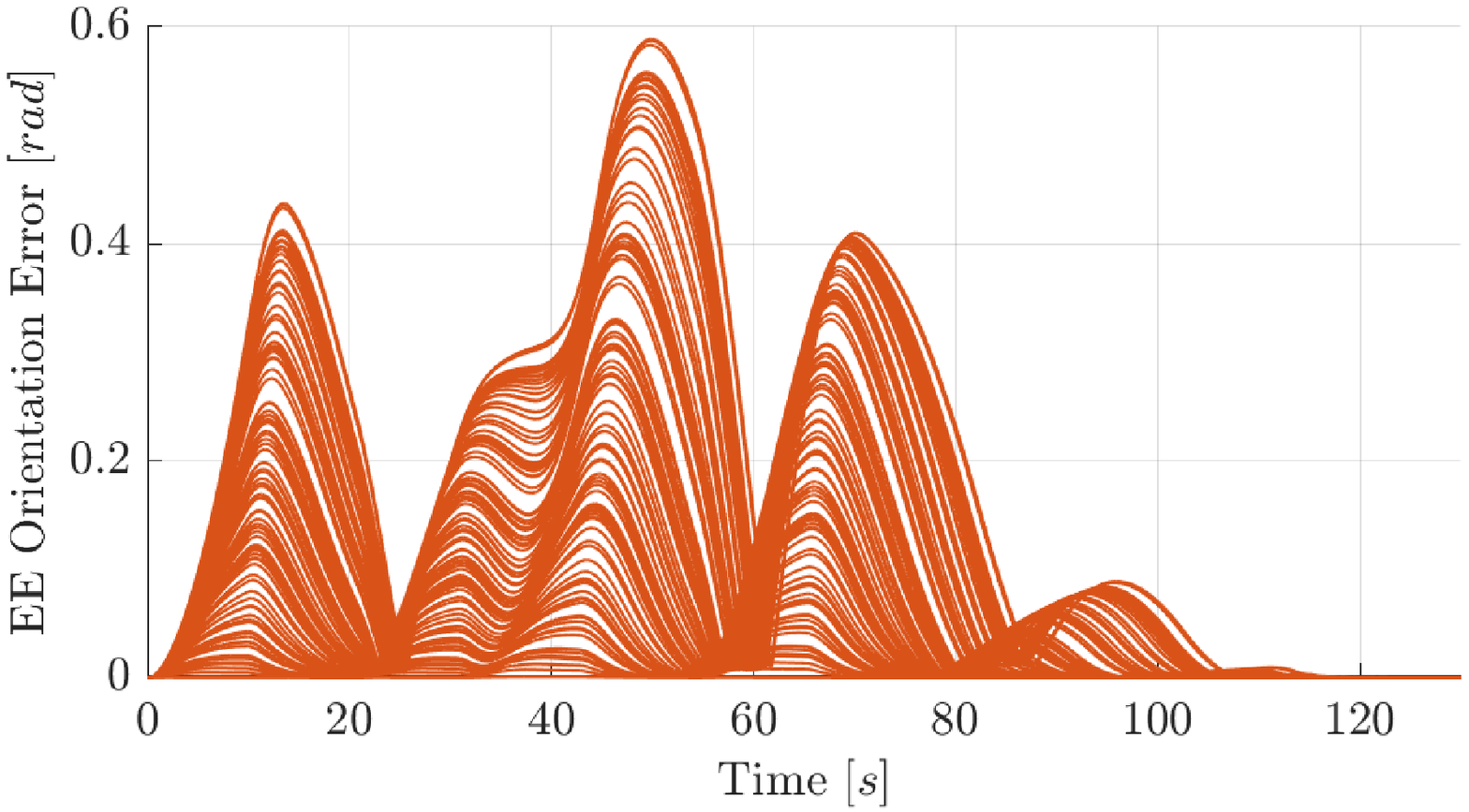}
\captionof{figure}{Prop. NTSMC controller position and attitude error norms of the EE during the MC simulations.}
\label{fig:S3}
\end{minipage}

\end{figure*}

One last important aspect that is worth studying is how the new controller deals with model uncertainties, and in particular what are the effects of unmodelled EE loads and BS uncertainties on the closed loop system performance. %
To this end, the response of control system is analyzed in 100 Monte Carlo (MC) simulations during which the masses, moment of inertia (MoI) and product of inertia (PoI) of the BS and of the EE are selected randomly. 
In particular, the mass of the BS and MoI are sampled uniformly over the $\pm 5 \%$ range w.r.t. their nominal values, and its PoI is selected over an uniform distribution ranging from $0$ to the $10 \%$ of the MoI nominal value instead. EE  mass, MoI and PoI are selected respectively over the following uniform distributions: $\mathcal{U}(0.1,100.1)\ [kg]$, $\mathcal{U}(50,150) \ [kgm^2]$, $\mathcal{U}(0,10)\ [kgm^2]$. 
The EE position and orientation error norms resulting from the MC simulations are reported in Fig.~\ref{fig:S3}: from this image, it is evident that the designed control system is able to keep the tracking error of the EE bounded and bring it to zero besides the presence of even strong uncertainties.}{


\section{Simulation and discussion}
\label{sec:sim}

\begin{figure*}
\centering
\hspace{10 pt}
\begin{subfigure}[]{0.25\textwidth}
\includegraphics[trim = 0 0 10 0, clip,width=0.9\columnwidth]{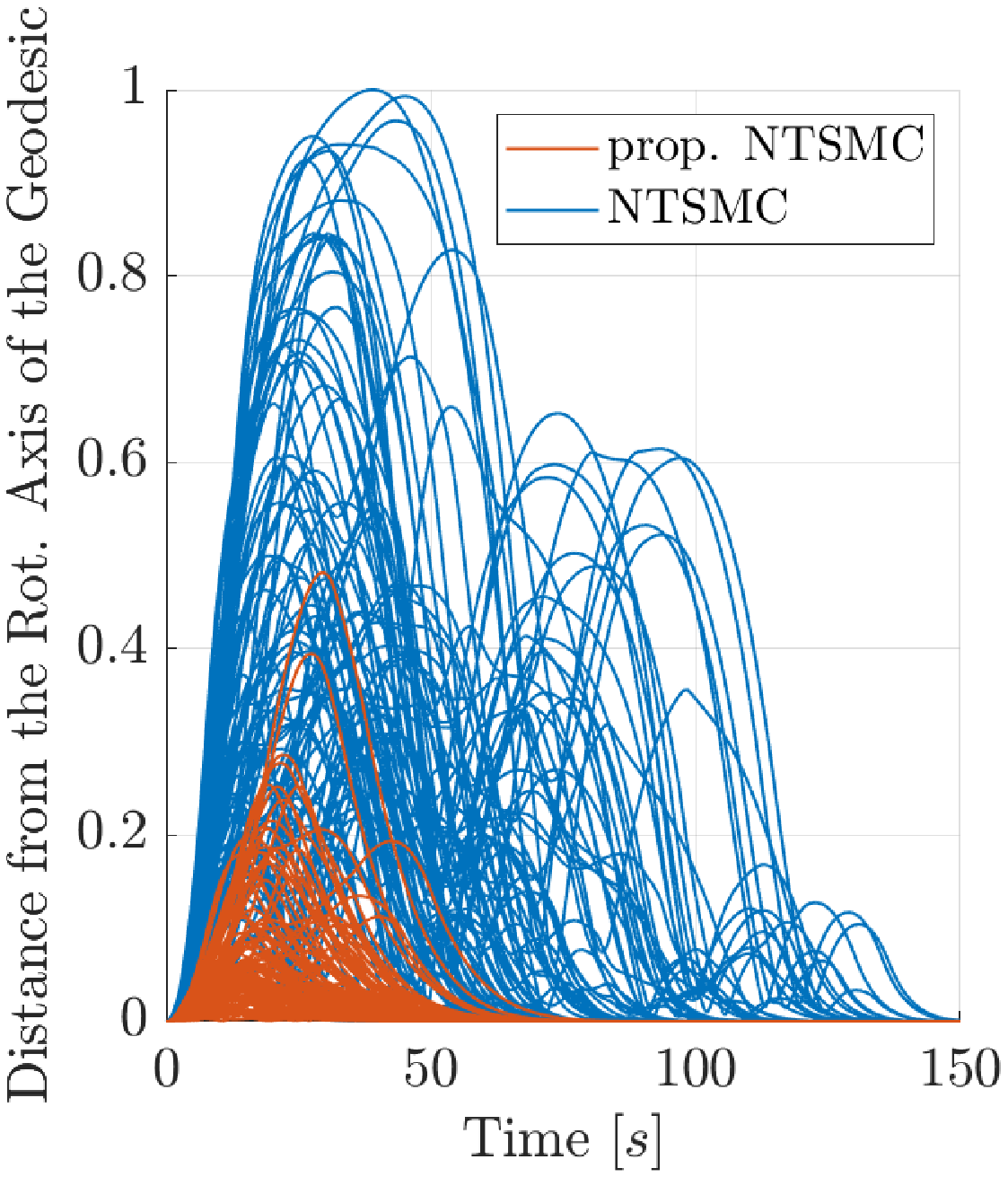}
\caption{MC campaign}
\label{fig:S1}
\end{subfigure}
\begin{subfigure}[]{0.36\textwidth}
\includegraphics[trim = 10 0 20 10, clip,width=0.9\columnwidth]{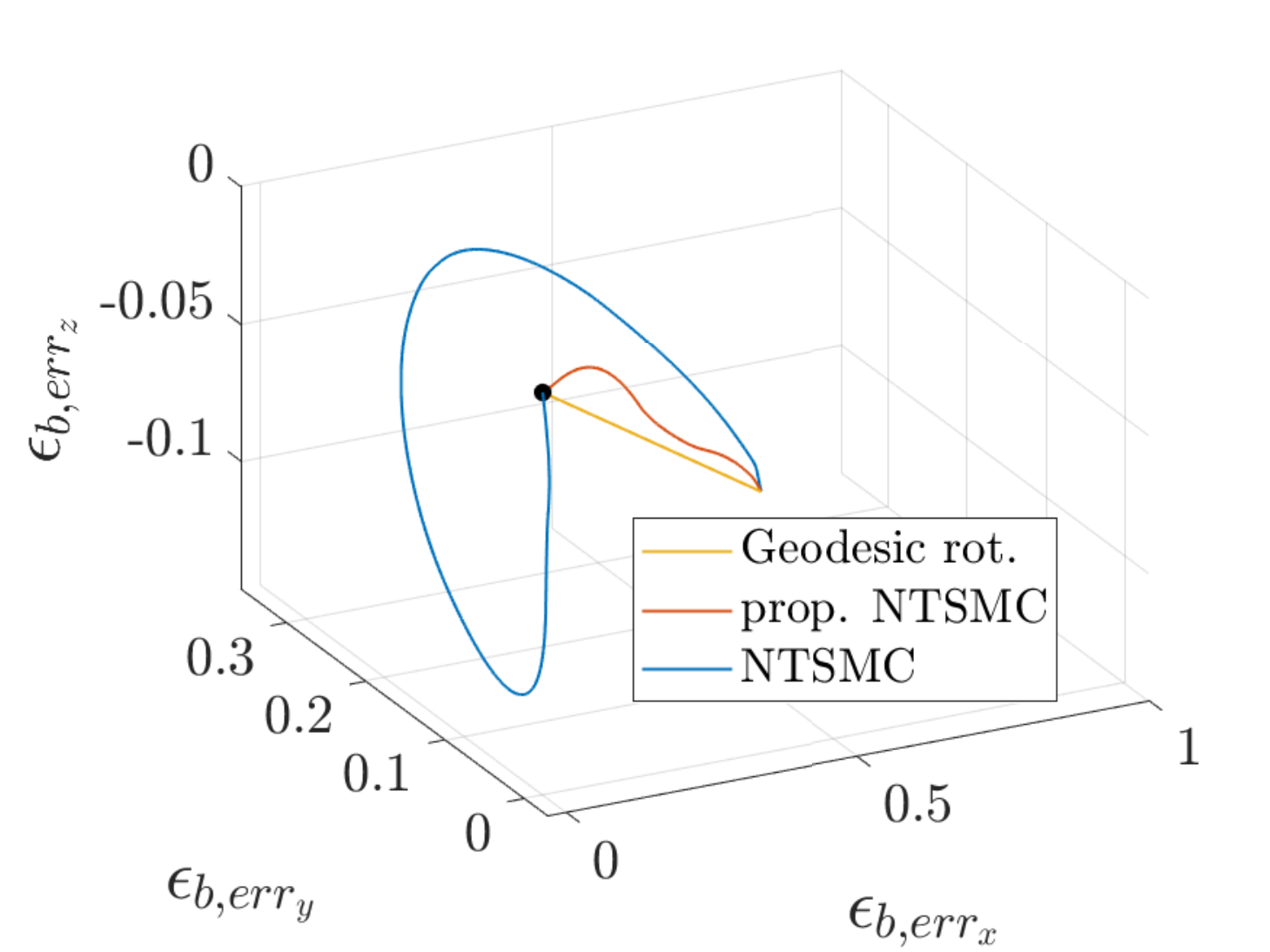}
\caption{Single MC run}
\label{fig:S2}
\end{subfigure}
\hspace{10 pt}
\begin{subfigure}[]{0.32\textwidth}
\includegraphics[trim = 30 162 45 140, clip,width=0.8\columnwidth]{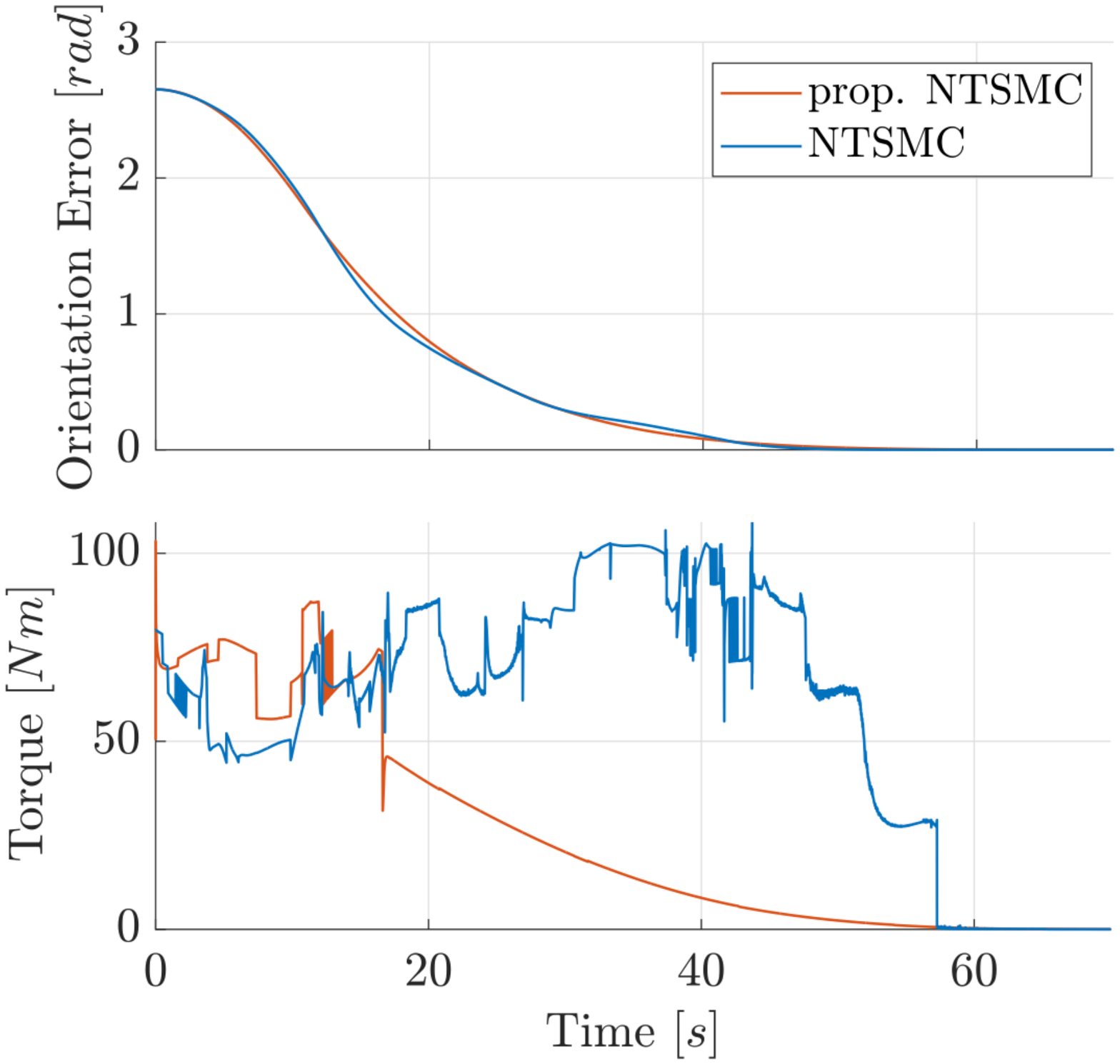}
\caption{Single MC run}
\label{fig:S3}
\end{subfigure}
\caption{Proposed NTSMC vs NTSMC\cite{smNTSMC} controllers. (a) Euclidean distances of the controllers instantaneous rotation axes w.r.t. that associated with the geodesic rotation during each MC run. (b) Evolution of the vector part of the error quaternion to zero. 
(c) BS attitude error and control torque norms.}
\vspace{-8pt}
\end{figure*}

\LSv{
\begin{figure*}
\begin{minipage}{0.3\linewidth}
\centering
\includegraphics[trim = 0 0 0 0, clip,width=0.8\textwidth]{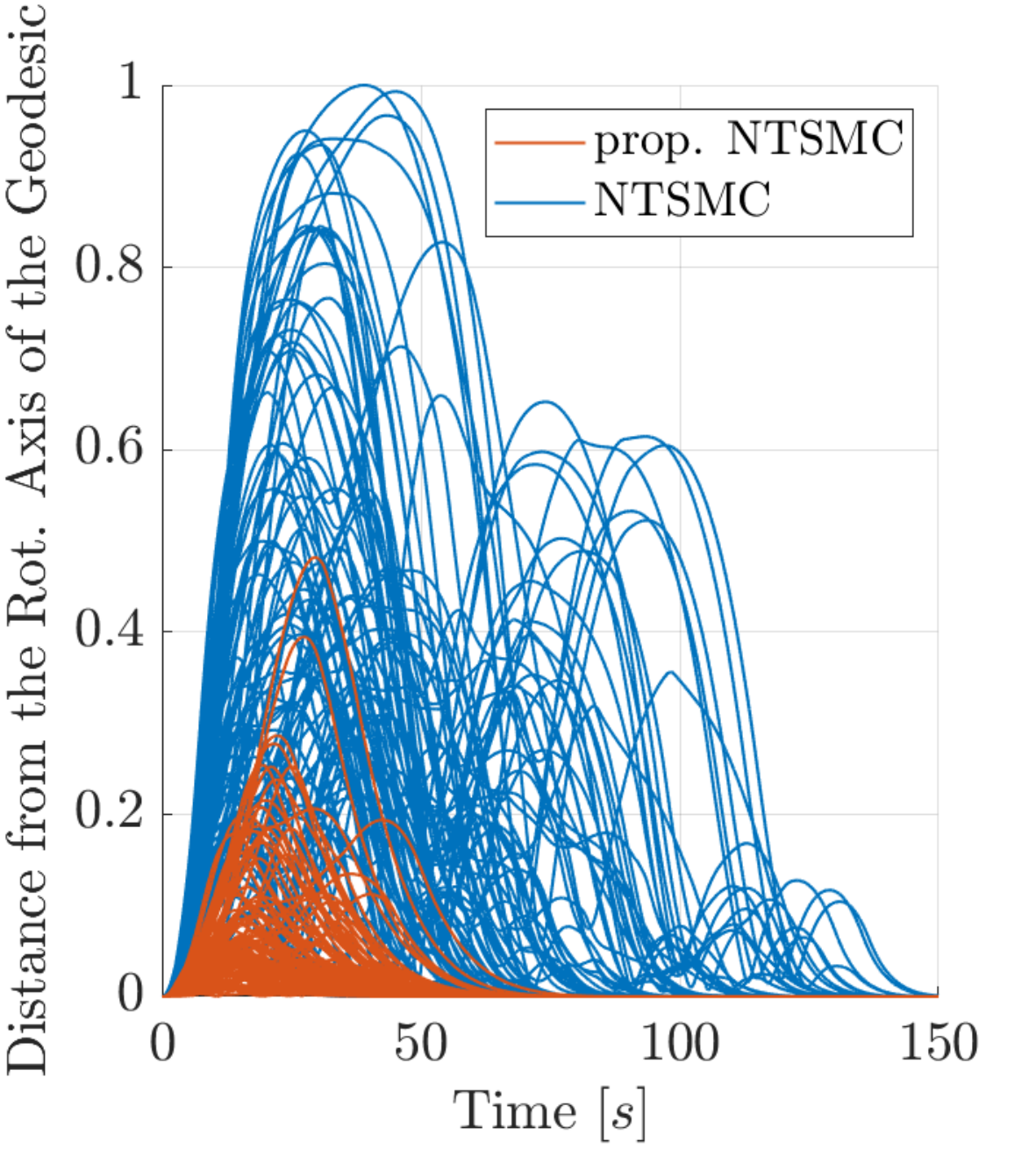}
\captionof{figure}{Euclidean distances of the two controllers instantaneous rotation axes w.r.t. the one associated with the geodesic rotation during each MC run.}
\label{fig:S1}
\end{minipage}
\hfill
\begin{minipage}{0.69\linewidth}
    \begin{minipage}{0.5\linewidth}
    \centering
    \includegraphics[trim = 0 0 20 10, clip,width=1\textwidth]{imgs/S2_rot}
    \end{minipage}
    \hfill
    \begin{minipage}{0.45\linewidth}
    \centering
    \includegraphics[trim = 22 0 45 0, clip,width=0.78\textwidth]{imgs/S2_o}
    \includegraphics[trim = 0 0 45 0, clip,width=0.8\textwidth]{imgs/S2_t}
    \end{minipage}
    \captionof{figure}{Proposed NTSMC vs NTSMC\cite{smNTSMC} controllers. Left: evolution of the vector part of the error quaternion. Given the axis-angle parametization of the orientation error, $\bm{\epsilon}_{b,e}$ can be interpreted as the path of the rotation axis scaled by half the sine of the rotation angle. For this reasons it gives a hint on the rotation direction and its magnitude. For reference, the geodesic rotation, i.e. the shortest one, is shown in yellow. Right: BS attitude error and control torque norms.}
    \label{fig:S2}
\end{minipage}
\end{figure*}
}
The effectiveness of the proposed quaternion based control solution is analyzed through a hundred Monte Carlo (MC) simulations: for each MC run, a random misalignment for the BS w.r.t. the reference trajectory is selected, by choosing three values in the interval $\pm \frac{4}{5} \pi [rad]$ for the $xyz$ Euler Angle triplet describing the (high) alignment error.  
The characteristics of the considered space manipulator are defined according to those of previously accomplished space missions. 
Performance comparisons are made w.r.t the recent NTSMC controller in~\cite{smNTSMC}, which is based on Euler Angles, and controllers are tuned so that they perform similarly in terms of tracking error performances (Tab. \ref{tab:contParam}). 

\begin{table}[b!]
    {\footnotesize
    \centering
    \caption{Controllers Parameters}
\begin{tabular}{|c|c|}
    \hline
    Controllers & Parameters\\
    \hline
    \multicolumn{1}{|c|}{\multirow{5}{*}{prop.  NTSMC}}    &  \multicolumn{1}{|c|}{\multirow{5}{*}{\shortstack{$p = 9$, $q=11$, $p_1=5$, $q_1=9$, $p_2=7$, $p_2=9$,\\ $\bm{\Gamma}_1 = 0.1  \bm{E}$, $\bm{\Gamma}_2 = 0.1 \bm{E}$, $\bm{K}_1= 10^{-2} \bm{E}$,\\ $\bm{K}_2= \text{diag}(0.1 \ \bm{1}_{1 \times 8},\ 0.2 \ \bm{1}_{1 \times 2}$,  $0.6 \ \bm{1}_{1 \times 4})$ \\ $\Phi= 10^{-3}$,  $\hat{\bm{K}}_{\bm{\delta}}(0) = 10^{-4}$, $\phi = 10^{-3}$ }}}\\ 
    & \\
    & \\
    & \\
    & \\
    \hline
    \multicolumn{1}{|c|}{\multirow{5}{*}{NTSMC \cite{smNTSMC}}}    &  \multicolumn{1}{|c|}{\multirow{5}{*}{\shortstack{$p = 9$, $q=11$, $p_2=5$, $q_2=9$, $p_3=7$, $p_3=9$,\\ $\bm{\beta}_1 = 0.1  \bm{E}$,  $k_0= 10^{-4}$, $k_1= 10^{-3}$,\\ $\bm{K}_2= \text{diag}(0.1 \ \bm{1}_{1 \times 8},\ 0.2 \ \bm{1}_{1 \times 2}$,  $0.6 \ \bm{1}_{1 \times 4})$\\ $\theta = 10^{-3}$, $c_0= 10^{-3}$}}}\\ 
    & \\
    & \\
    & \\
    & \\
    \hline
    \end{tabular}
    \label{tab:contParam}
}
\end{table}
}

The geodesic path between two points on a sphere, i.e. the shortest curve connecting the two, is the intersection of a plane passing through its center and connecting the two points. The rotation that moves one point onto the other along the geodesic can be represented using the axis-angle convention, and in such case, the axis coincides with the unitary vector normal to the above-mentioned plane that passes through the center of the sphere. Therefore, during the regulation task, the closer the instantaneous axis of rotation is to the axis associated to the geodesic the closer the followed path is to the shortest one. 
The Euclidean distances of the two controllers instantaneous rotation axes w.r.t. that associated with the geodesic during the MC campaign are reported in Fig.~\ref{fig:S1}. It is clear that the proposed solution stays closer to the shortest rotation w.r.t. Euler Angles based one.

Analyzing now a single MC run, it is possible to assess the performance of the controllers more in depth. 
Given the axis-angle parametization of the orientation error, the vector part of the quaternion error $\bm{\epsilon}_{b,e}$ can be interpreted as the path of the rotation axis scaled by half the sine of the rotation angle. In summary, it gives an indication on the rotation direction and its magnitude.
Indeed, looking at $\bm{\epsilon}_{b,e}$, shown in Fig.~\ref{fig:S2}, it can be noticed that in the case of the proposed controller such error practically maintains its direction and scales its magnitude only: 
this means that the rotation axis is kept almost the same during the attitude regulation. Also, it results to stay closer to the geodesic.
Finally, from Fig.~\ref{fig:S3}, it is clear that even if the convergence time of the attitude error norm is almost the same, the integral of torque norm required for the attitude correction is much lower when using the quaternion-based solution and this results in better fuel efficiency. On average across the MC campaign, the integral of torque norm is reduced by $62.78 \%$. Such an improvement can be linked to the fact that the proposed controller better captures the topology of $\mathbb{S}^3$. 

Additional simulations are reported in the Appendix, showing the robustness of the approach w.r.t. large or singular angular displacement and disturbances in the estimation/measurement of the manipulated mass and moment of inertia, with bounded errors converging to zero.

\vspace{-0pt}
\section{Conclusion}
\label{sec:conc}
In this work, a dynamic model for a space manipulator is derived using Lagrangian formalism. Then an NTSMC based on quaternions is proposed to control the system. Chattering problems are mitigated with a boundary layer thickness approach and an adaptive estimation of the disturbances and uncertainty components upper-bounds is proposed. 

The stability and nominal performance of the designed controller are first theoretically proven using Lyapunov theory, and then assessed in terms of efficiency and effective fuel consumption with an MC simulative approach. 
 
\bibliographystyle{IEEEtran}
\bibliography{bibliography}

\newpage
\renewcommand\thefigure{A.\arabic{figure}}    
\setcounter{figure}{0}    

\appendix 
\subsection{System scenario}
\begin{figure*}[!t]
    \centering
    \begin{minipage}{0.37\linewidth}
    \centering
    \includegraphics[width=1\textwidth]{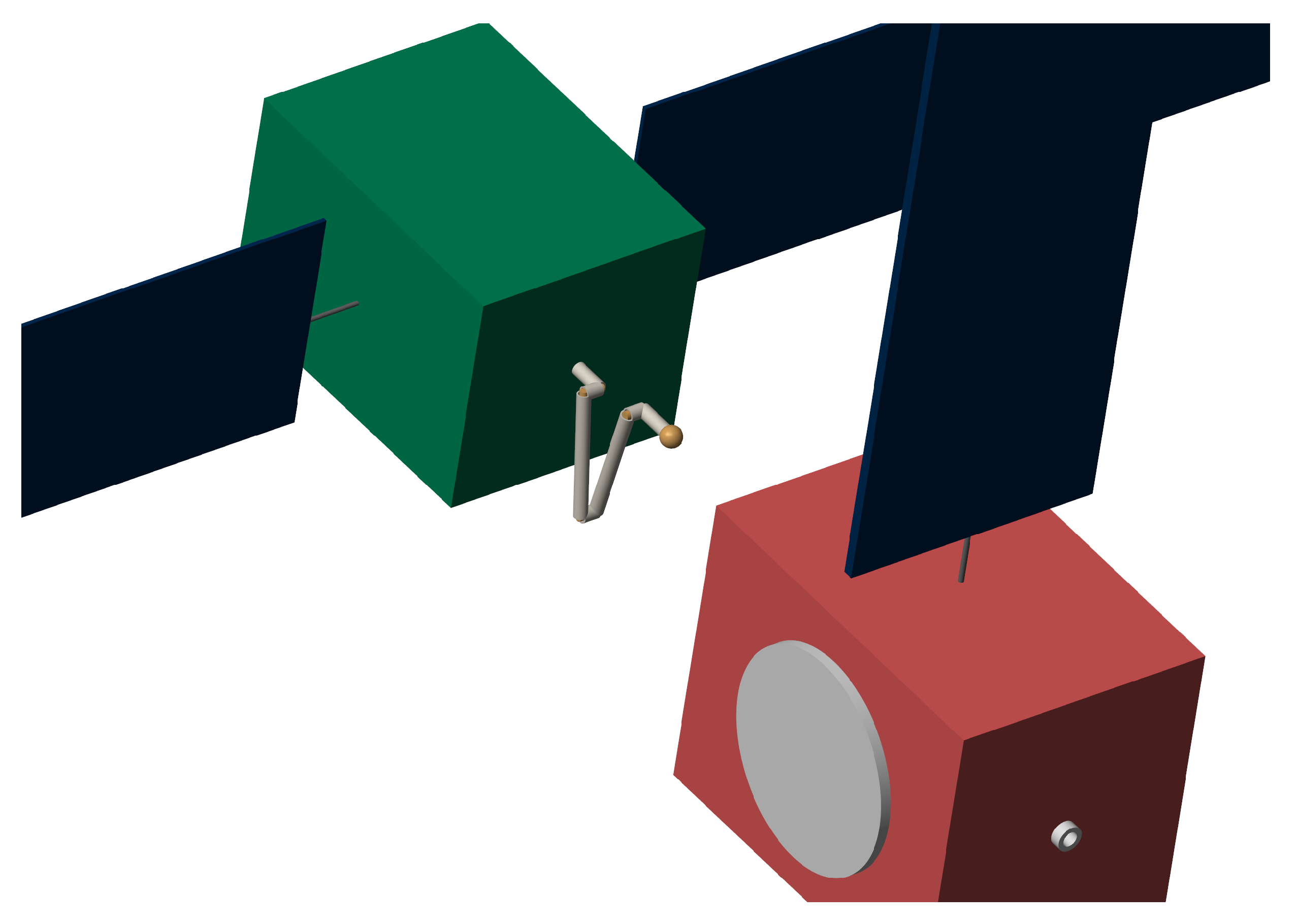}
    \end{minipage}
    \hspace{2pt}
    \begin{minipage}{0.61\linewidth}
    \centering
    \includegraphics[width=1\textwidth]{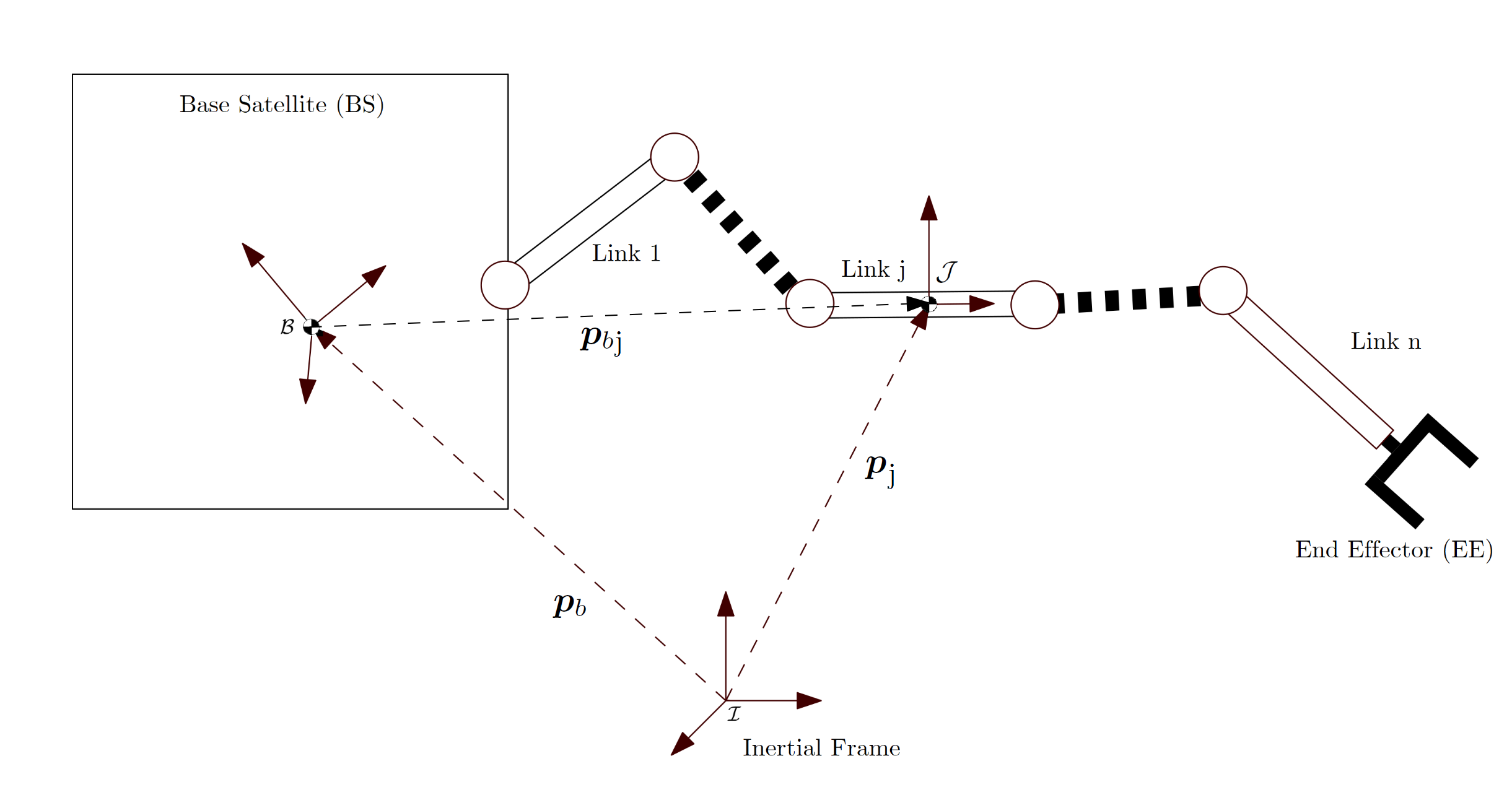}
    \end{minipage}
       \caption{System scenario. Rendering of the base satellite + robotic manipulator system.}
    \label{fig:SM_scenario}
\end{figure*}

The system studied in this work is shown in Fig.~\ref{fig:SM_scenario}, where a rendering of the scenario is provided on the left, and a more technical scheme is reported on the right. 
In particular, in the former, the Base Satellite (BS) is depicted in green, and the arm (in grey) with the End Effector (EE) (in yellow) is acting to capture a load (red satellite).
In the scheme on the right, instead, $n$ links connect the BS with the EE. The inertial frame $\cal I$, the body frame $\cal B$, and the link frame $\cal J$ are placed and the position of the $j^{th}$ link w.r.t. $\cal B$ and $\cal I$ is highlighted. 

\subsection{Simulations with critical scenarios}
Two examples where the Euler-based controller in~\cite{smNTSMC} performs poorly w.r.t. the proposed one are hereafter shown.

It is widely known that Euler Angles representations suffer from singularities. For example, when using the $xyz$ triplet it is not possible to reconstruct the angles of an elementary rotation from a given rotation matrix if a rotation of $\pm \pi/2 \ [rad]$ w.r.t. the $y$ axis is involved. In the same point also the conversion from angular velocities to Euler Angles derivatives is problematic. 

\begin{figure*}[!h]
    \begin{minipage}{0.332\linewidth}
        \centering
        \includegraphics[width=1\textwidth]{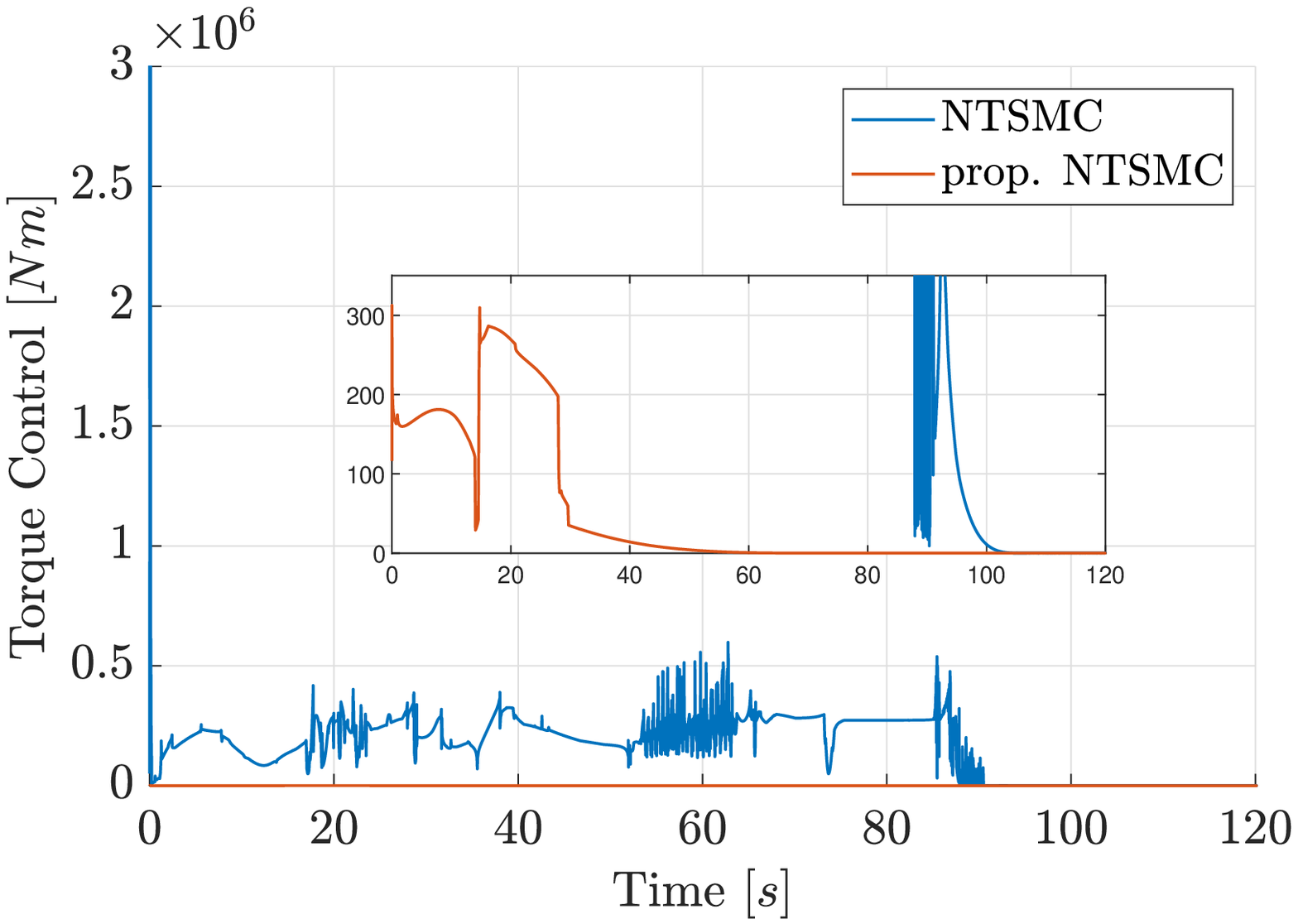}
    \end{minipage}
    \begin{minipage}{0.332\linewidth}
        \centering
        \includegraphics[width=1\textwidth]{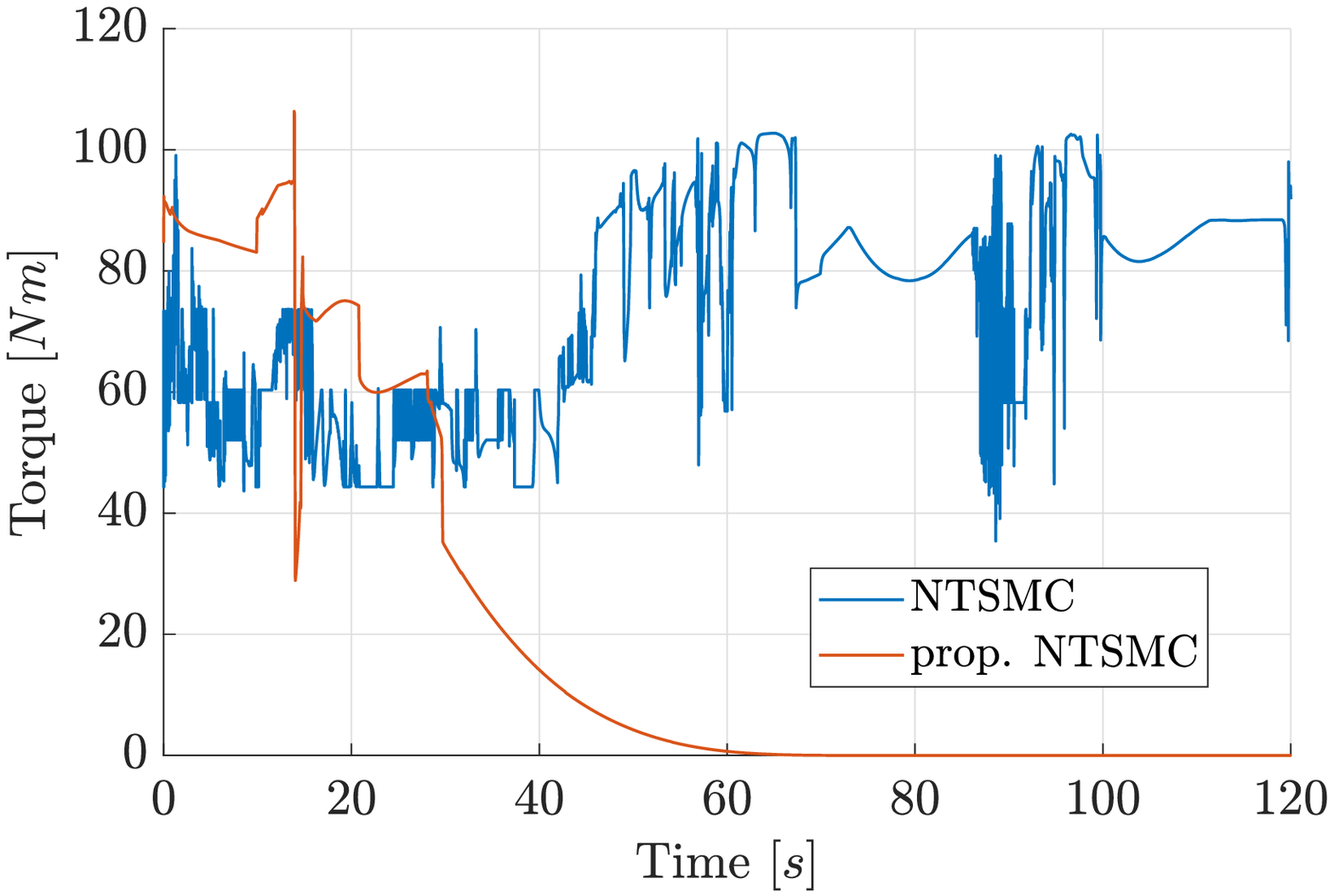}
    \end{minipage}    
    \begin{minipage}{0.332\linewidth}
        \centering
        \includegraphics[width=1\textwidth]{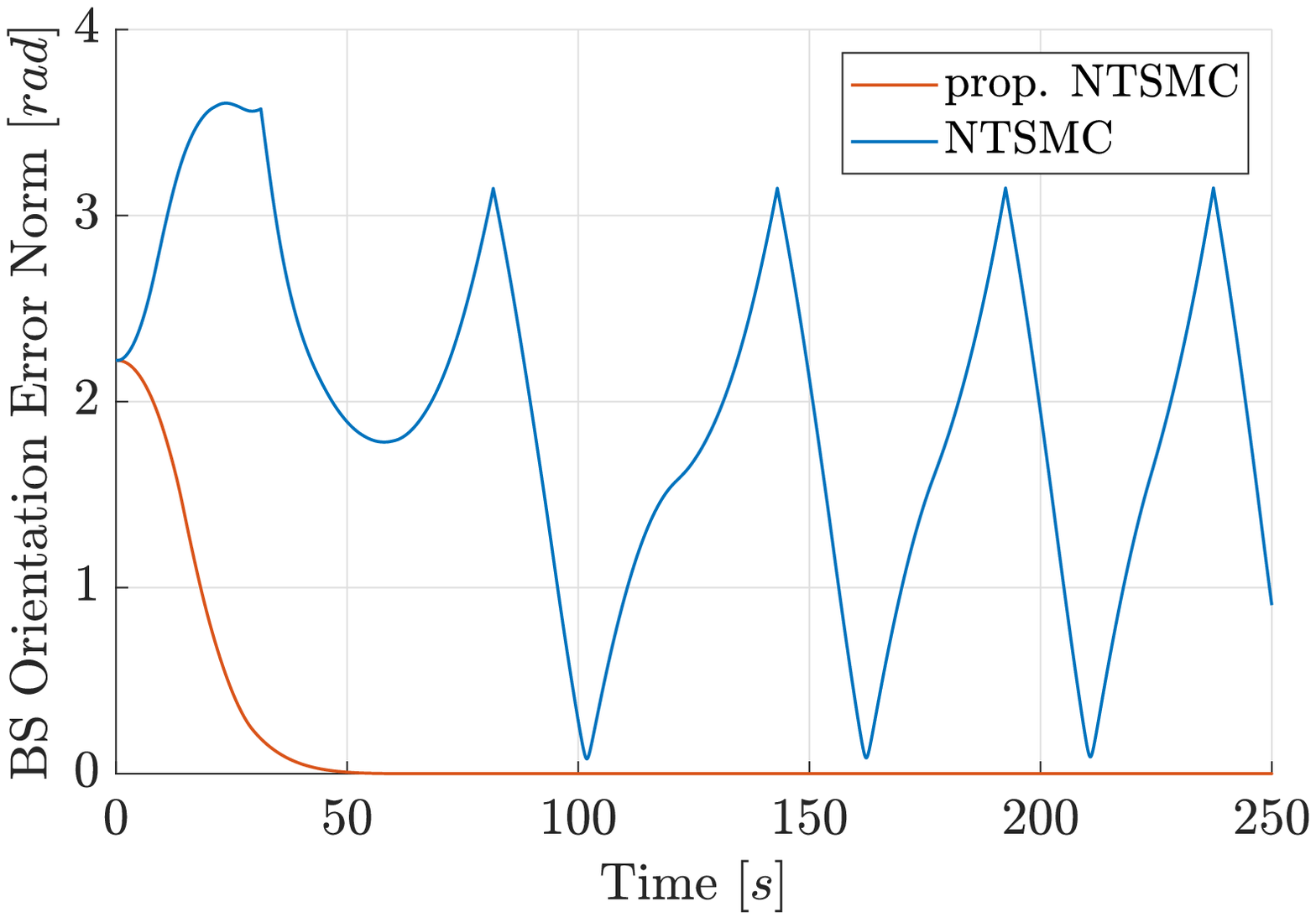}
    \end{minipage}
        \caption{Proposed NTSMC vs NTSMC\cite{smNTSMC} controllers response for a BS displacement of $[-5/4\pi,\pi/2,-5/4\pi]_{xyz} \ [rad]]$ w.r.t. the reference trajectory.}
        \label{fig:F_sing}
\end{figure*}

\begin{figure*}[!h]
    \begin{minipage}{0.332\linewidth}
        \centering
        \includegraphics[width=1\textwidth]{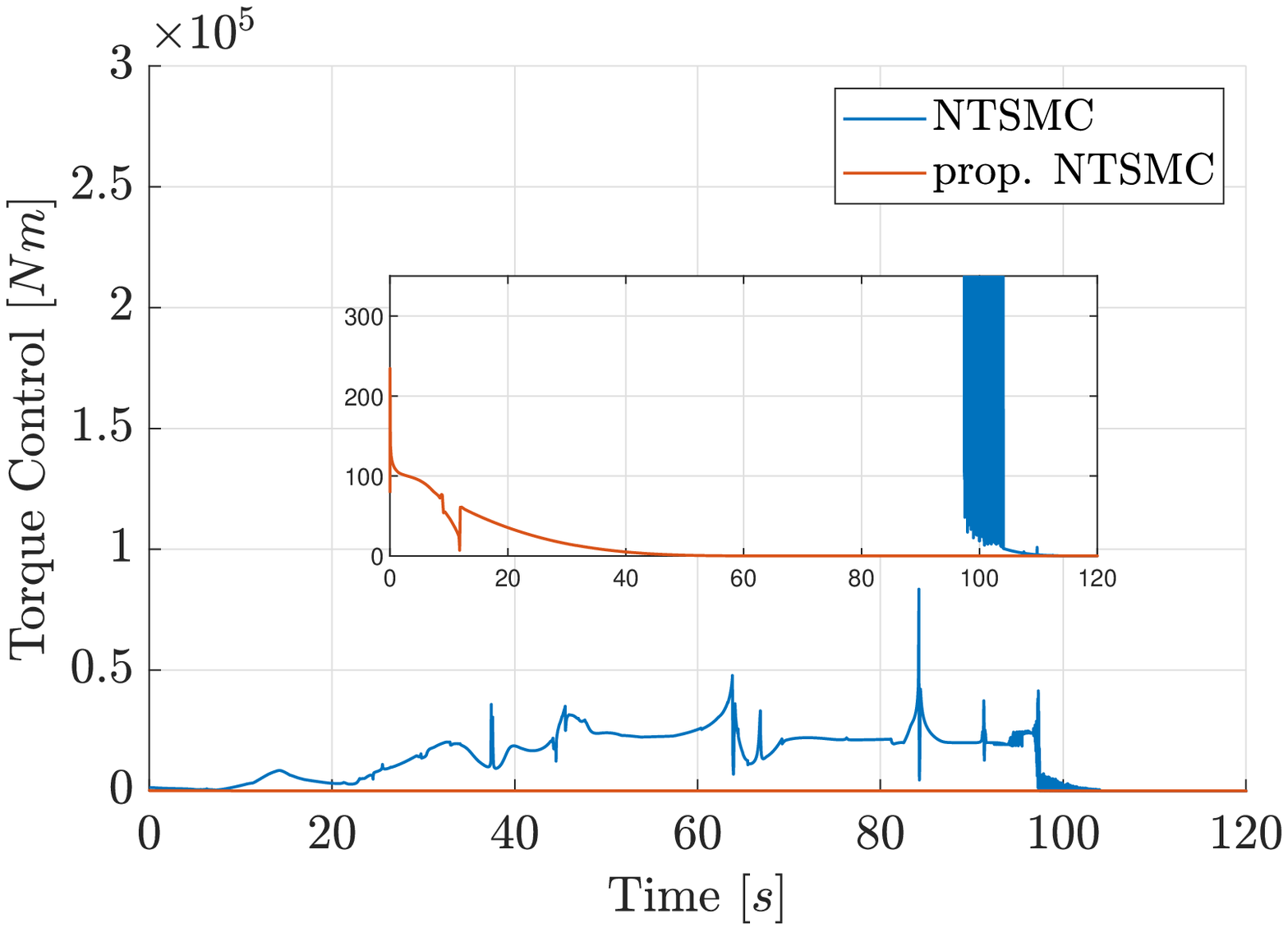}
    \end{minipage}
    \begin{minipage}{0.332\linewidth}
        \centering
        \includegraphics[width=1\textwidth]{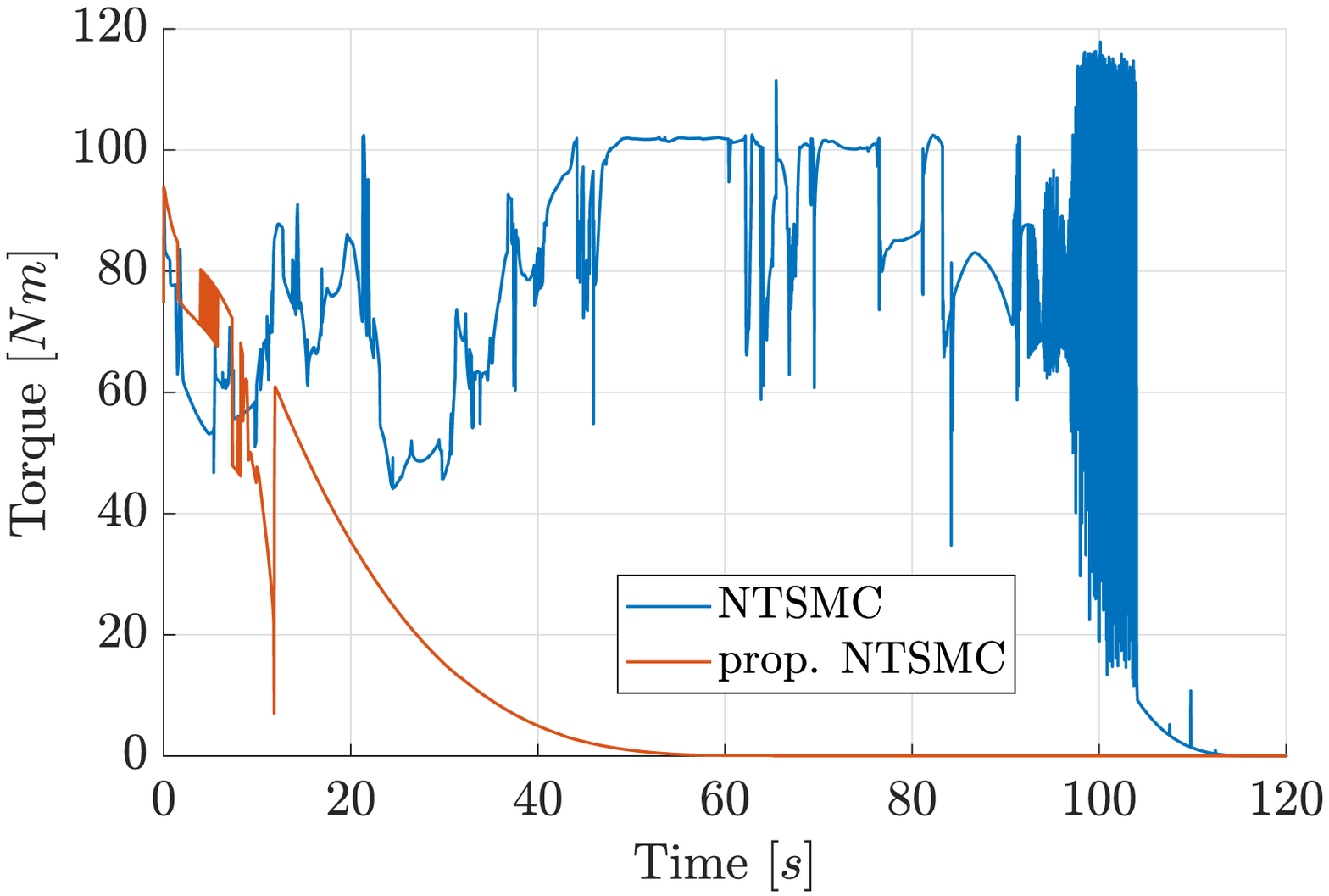}
    \end{minipage}    
    \begin{minipage}{0.332\linewidth}
        \centering
        \includegraphics[width=1\textwidth]{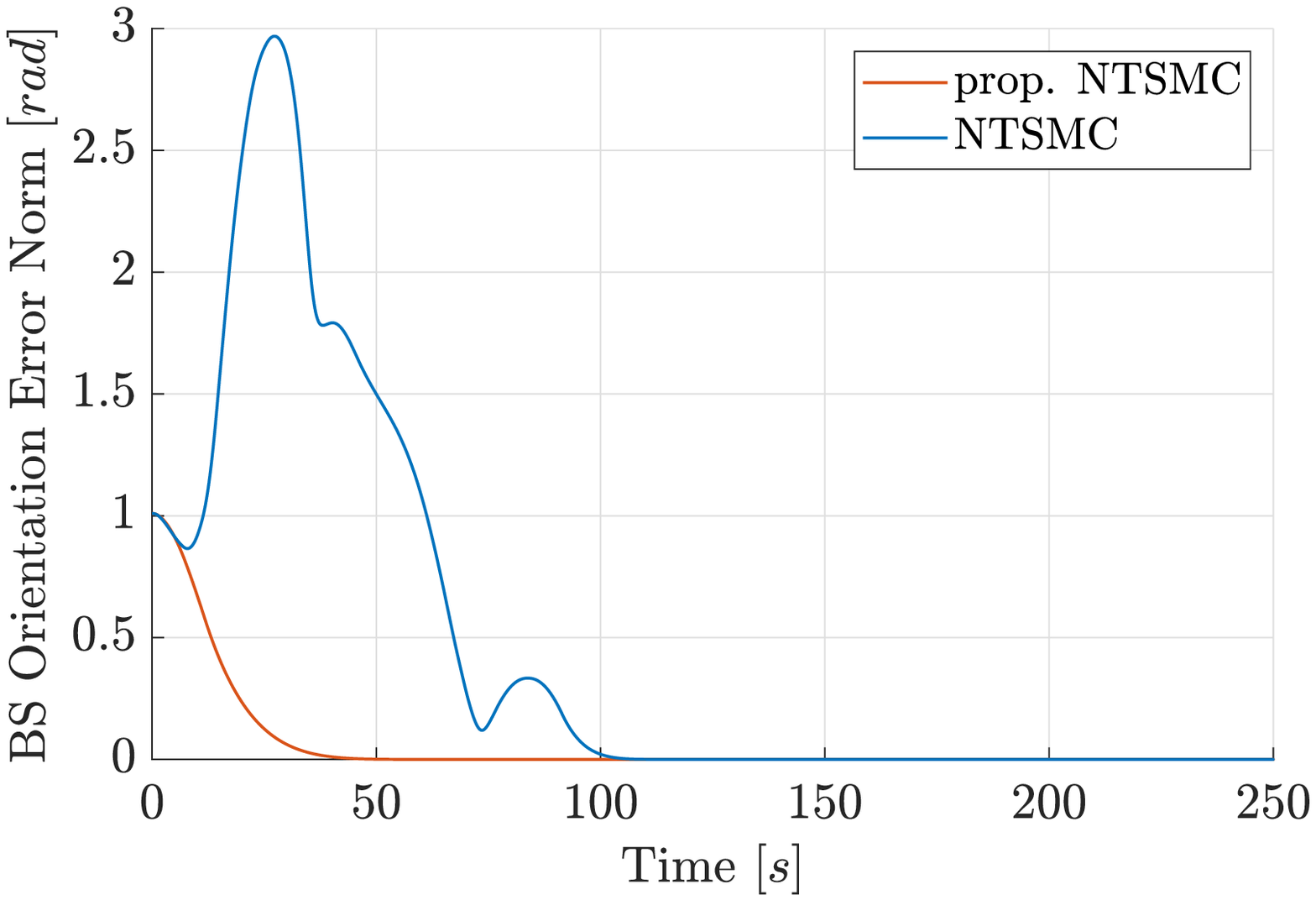}
    \end{minipage}
        \caption{Proposed NTSMC vs NTSMC\cite{smNTSMC} controllers response for a BS displacement of $[-7/4\pi, \pi/6,-5/4\pi]_{xyz} \ [rad]$ w.r.t. the reference trajectory.}
        \label{fig:F_larg}
\end{figure*}

Unfortunately, these issues are implicitly embedded in Euler Angle based controllers and require ad hoc heuristics in order to avoid unbounded control input and instability problems. In addition, it is hard to tune an Euler Angles based controller gains to be effective in both small and large attitude displacements situations. Instead, these problems are not present in the case of quaternion based controllers. Two remarkable examples are shown in Fig.~\ref{fig:F_sing} and Fig.~\ref{fig:F_larg}: in the first one a near-singularity configuration is tested and in the second one a large attitude displacement is considered. 
The three panels of each figure report the norm of the requested control input before saturation (left), the norm of the applied torque (center), and the evolution of the orientation error in norm (right). In this respect, some observations are in order: first, we note that the saturation is applied to the different actuators and therefore to the different directions along which they are acting, hence the relation between the first two plots is not trivial.
Secondly, the time scale of the error behavior has been extended in order to better highlight the different behavior of the controllers towards or after convergence.

In both the reported cases of Fig.~\ref{fig:F_sing} and Fig.~\ref{fig:F_larg}, the unsaturated control input of the Euler based controller is large and leads to the above-mentioned undesired behavior, appearing as high input requests, oscillatory behaviors, and long convergences, if any. 

\subsection{Performance in presence of disturbances}

With the aim of studying the effects of unmodelled EE loads and BS uncertainties, the trajectory tracking response of the control system is analyzed with the support of 100 Monte Carlo (MC) simulations during which the masses, moment of inertia (MoI), and product of inertia (PoI) of the BS and of the EE are selected randomly. Specifically, the mass of the BS and MoI are sampled uniformly over the $\pm 5 \%$ range w.r.t. their nominal values and its PoI is selected over a uniform distribution ranging from $0$ to the $10 \%$ of the MoI nominal value instead. EE  mass, MoI and PoI are selected respectively over the following uniform distributions: $\mathcal{U}(0.1,100.1)\ [kg]$, $\mathcal{U}(50,150) \ [kgm^2]$, $\mathcal{U}(0,10)\ [kgm^2]$. The trajectory chosen for this test lasts about $57 \ [s]$ and consists of a diagonal movement of the BS while keeping the robotic arm fixed followed by another diagonal movement of the EE with the BS pose fixed.

\begin{figure*}[!t]
    \centering
    \includegraphics[width=0.495\textwidth]{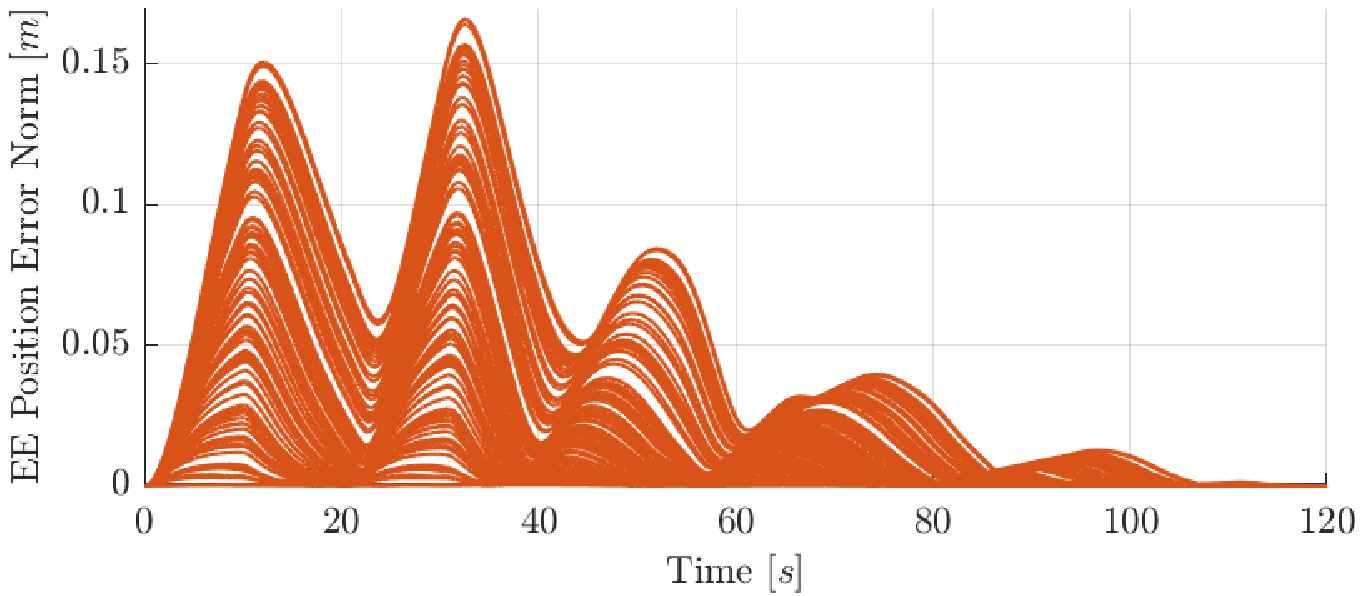}
    \includegraphics[width=0.495\textwidth]{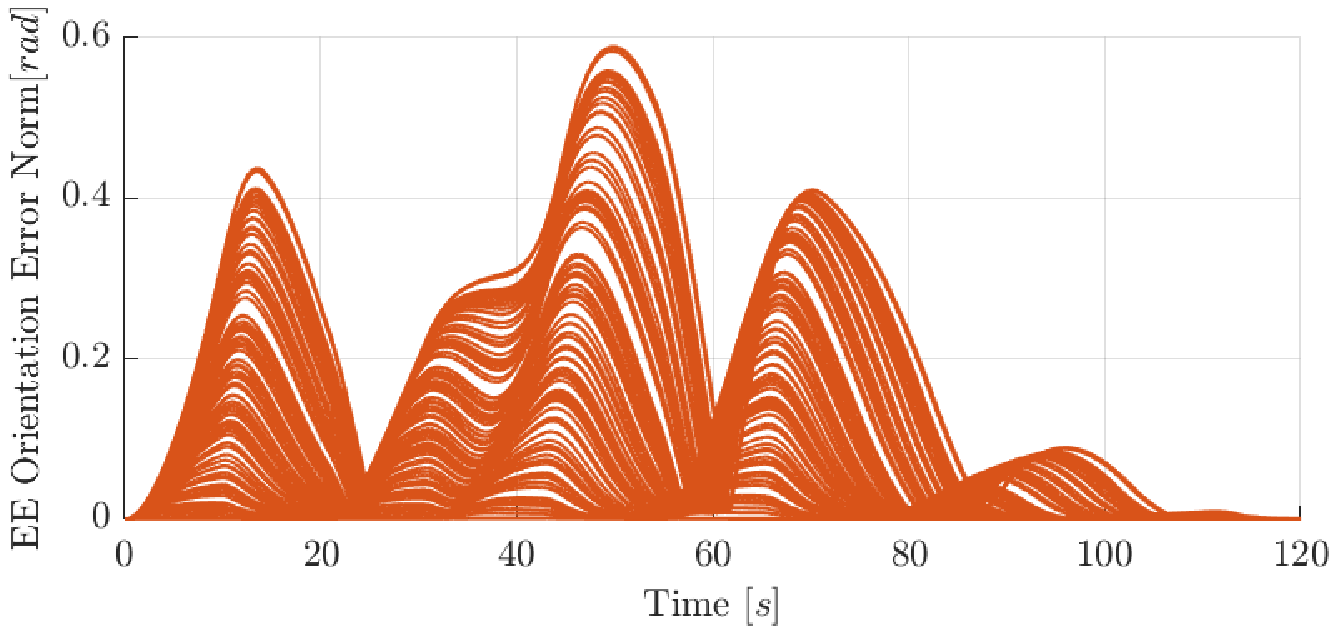}
    \caption{MC simulations: proposed NTSMC in presence of disturbances}
    \label{fig:MC_dist}
\end{figure*}

The EE position and orientation error norms resulting from the MC simulations are reported in Fig.~\ref{fig:MC_dist}: from this image, it is evident that the designed control system is able to keep the tracking error of the EE bounded and bring it to zero besides the presence of uncertainties in a reasonable time. More in detail, in terms of EE position an error lower than $1 \ [mm]$ is reached on average in $87.80 \ [s]$ with a sample variance of $13.27 \ [s]$ and in terms of orientation an error lower than  $0.01 \ [deg]$ is obtained on average in $100.63 \ [s]$ with a sample variance of $9.57 \ [s]$.

{}



\end{document}